\documentclass[aps, prb, amsmath, twocolumn, longbibliography, superscriptaddress, footinbib, 10pt]{revtex4-2}

\usepackage{ifthen}
\usepackage{amssymb, amsmath,amsfonts}
\usepackage[per-mode=symbol]{siunitx} 
\usepackage[final]{microtype} % paper will look much better: only work in final mode, not draft
\usepackage{setspace}
\usepackage{xcolor}
\usepackage[utf8]{inputenc}
\usepackage[l2tabu, orthodox]{nag}

\usepackage{physics}
\usepackage{mathtools}% http://ctan.org/pkg/mathtools
\usepackage{bm}
\usepackage{upgreek}
\usepackage{booktabs} % The booktabs package creates much nicer looking tables than the standard latex tables
\usepackage[version=4]{mhchem}

\usepackage{array} % its ability to create custom columns is invaluable for formatting tabular material on a per-column basis.
\usepackage{adjustbox}
\usepackage[flushleft]{threeparttable}
\usepackage{threeparttablex} % for "ThreePartTable" environment
\usepackage{comment}
\usepackage{calc}

\usepackage{lmodern}      % Police de caractères plus complète et généralement indistinguable visuellement de la police standard de LaTeX (Computer Modern).
\usepackage[T1]{fontenc}  % Bon encodage des caractères pour qu'Acrobat Reader reconnaisse les accents et les ligatures telles que ffi.

\usepackage{graphicx}
\usepackage{flafter,placeins}
\usepackage{float}
\usepackage{color,soulutf8,colortbl,xspace}
\usepackage[skins]{tcolorbox}
\usepackage{multirow}
\usepackage{makecell}
\makeatletter
\let\ps@plain=\ps@pagenumber
\makeatother

\usepackage[acronym]{glossaries}
\usepackage[colorlinks=true,linkcolor=blue, citecolor=blue, urlcolor=blue]{hyperref}
\usepackage[all]{hypcap}

\newcommand{\polydept}{Department of Engineering Physics, \'Ecole Polytechnique de Montr\'eal, C.P. 6079, Succ. Centre-Ville, Montr\'eal, Qu\'ebec, Canada H3C 3A7}

\newcommand{\rajshahidept}{Solar Energy Laboratory, University of Rajshahi 6205, Bangladesh}

\newcommand{\bostondept}{Department of Engineering, University of Massachusetts Boston, Boston, MA 02125, USA}

\newcommand{\GeSn}[2]{Ge$_{#1}$Sn$_{#2}$}

\newcommand{\quot}[1]{``#1''} % quote (ou gillemets)
\newcommand{\pin}{$p$-$i$-$n$}
\newcommand{\pnn}{$p$-$n$-$n$}
\newcommand{\Voc}{$V_{\text{OC}}$}
\newcommand{\Jsc}{$J_{\text{SC}}$}
\newcommand{\comments}[1]{} %defines a command that does nothing with the input (effectively commenting it out!)

\newcommand{\fref}[2]{\autoref{#1}\textcolor{blue}{#2}}
\newcommand{\tnoteMod}[1]{\tnote{\textcolor{blue}{#1}}}

\newcommand{\paperSection}[2][normal]{
    \ifthenelse{\equal{#1}{normal}}{
        \medskip\noindent{\textbf{#2}}\newline
    }{
        \noindent\textbf{#2}\newline
    }
}

\newcommand{\ac}[1]{\gls*{#1}}
\newcommand{\acpl}[1]{\glspl*{#1}}
\graphicspath{{./figures}{./figures/downsized_image/}}
\hypersetup{
  pdftitle={Group IV Mid-Infrared Thermophotovoltaic Cells on Silicon},
  pdfauthor={Gérard Daligou,  Richard Soref, Anis Attiaoui,  Jaker Hossain, Mahmoud R. M. Atalla, Patrick Del Vecchio, Oussama Moutanabbir},
  pdfkeywords={Mid-infrared, Germanium tin semiconductors,  Thermophotovoltaics, Photovoltaics, Waste heat recovery},
  bookmarksnumbered,
  pdfstartview={FitV},
  linktoc=all
}

\makeglossaries
\begin{document}

% acronyms
% \input{acronymDefinition.tex}
\newacronym{PBA}{PBA}{parabolic band approximation}
\newacronym{SRH}{SRH}{Shockley-Read-Hall}
\newacronym{OTMM}{OTMM}{optical transfer matrix method}
\newacronym{GTMM}{GTMM}{generalized transfer matrix method}
\newacronym{PCE}{PCE}{power conversion efficiency}
\newacronym{FSI}{FSI}{front-side illumination}
\newacronym{BSI}{BSI}{back-side illumination}
\newacronym{FF}{FF}{fill factor}
\newacronym{IQE}{IQE}{internal quantum efficiency}
\newacronym{TPV}{TPV}{thermo-photovoltaic}
\newacronym{ICPV}{ICPV}{interband cascade photovoltaic cells}
\newacronym[plural=EHPs, firstplural=electron-hole pairs (EHPs)]{EHP}{EHP}{electron-hole pair}
\newacronym{BSR}{BSR}{back surface reflector}
\newacronym{ICL}{ICL}{interband cascade laser}
% -------------------------------------------------------

% \title{GeSn-based thermophotovoltaic (TPV) devices}
\title{Group IV Mid-Infrared Thermophotovoltaic Cells on Silicon}
% \date{\today}

\author{G\'erard Daligou}
\affiliation{\polydept{}}

\author{Richard Soref}
\affiliation{\bostondept{}}

\author{Anis Attiaoui}
\affiliation{\polydept{}}

\author{Jaker Hossain}
\affiliation{\rajshahidept{}}

\author{Mahmoud R. M. Atalla}
\affiliation{\polydept{}}

\author{Patrick Del Vecchio}
\affiliation{\polydept{}}

\author{Oussama Moutanabbir}
\email{oussama.moutanabbir@polymtl.ca}
\affiliation{\polydept{}}

\begin{abstract}
\medskip
Compound semiconductors have been the predominant building blocks for the current mid-infrared thermophotovoltaic devices relevant to sub-\SI{2000}{\kelvin} heat conversion and power beaming. However, the prohibitively high cost associated with these technologies limits their broad adoption. Herein, to alleviate this challenge we introduce an all-group IV mid-infrared cell consisting of \GeSn{}{} alloy directly on a silicon wafer. This emerging class of semiconductors provides strain and composition as degrees of freedom to control the bandgap energy thus covering the entire mid-infrared range. The proposed thermophotovoltaic device is composed of a fully relaxed \GeSn{0.83}{0.17} double heterostructure corresponding to a bandgap energy of \SI{0.29}{\electronvolt}. A theoretical framework is derived to evaluate cell performance under high injection. The black-body radiation absorption is investigated using the generalized transfer matrix method thereby considering the mixed coherent/incoherent layer stacking. Moreover, the intrinsic recombination mechanisms and their importance in a narrow bandgap semiconductor were also taken into account. In this regard, the parabolic band approximation and Fermi’s golden rule were combined for an accurate estimation of the radiative recombination rate. Based on these analyses, power conversion efficiencies of up to 9\% are predicted for \GeSn{0.83}{0.17} thermophotovoltaic cells under black-body radiation at temperatures in the 500-$\SI{1500}{\kelvin}$ range. A slight improvement in the efficiency is observed under the frontside illumination but vanishes below $\SI{800}{\kelvin}$, while the use of a backside reflector improves the efficiency across the investigated black-body temperature range. The effects of the heterostructure thickness, surface recombination velocity, and carrier lifetime are also elucidated and discussed.
\end{abstract}

\maketitle

% load the introduction
%auto-ignore
\section{INTRODUCTION}\label{sec:intro}

% 1. The importance of TPV in the broad landscape of carbon-free energy conversion technologies
    
% 2. The relevance of MIR TPV. Why we need them? And what for? What are the applications that can only be served by MIR TPV (lower temperature energy waste).
    
% 3. The current state-of-the-art and challenges: InAs- and GaSb-based TPV, scalability, manufacturability, and cost-effectiveness.

\begin{figure*}[htb]
    \centering
    \includegraphics[width=\textwidth,keepaspectratio]{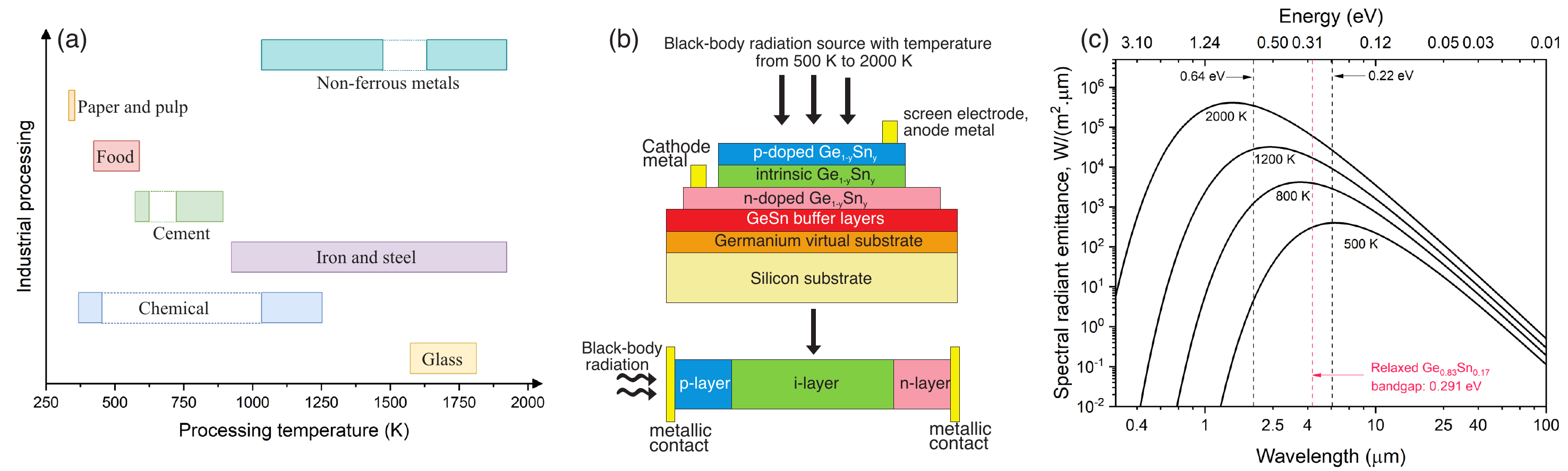}
    \caption{(a) Processing temperatures used in major manufacturing industries. The data presented in this chart, were mainly extracted from various U.S.~Department of Energy reports \cite{johnsonWasteHeatRecovery2008, thekdiIndustrialWasteHeat2015} and \cite{Papapetrou2022319}. Note that, within the same industry, the temperature range depends on the energy source used to fuel the foundries as well as on the nature of the product developed. For instance, in the industry of non-ferrous metals producing nickel requires a much higher temperature (up to $\SI{1923}{\kelvin}$) as compared to copper (up to $\SI{1363}{\kelvin}$) or aluminum (up to $\SI{1473}{\kelvin}$). (b) Schematic illustration of the \GeSn{1-y}{y} \ac{TPV} cell. (c) Evolution of the spectral radiant emittance with wavelength and temperature. The black dashed line indicates the bandgap energy of relaxed \GeSn{0.95}{0.05} and \GeSn{0.80}{0.20} semiconductors. At $\SI{500}{\kelvin}$, assuming no sub-bandgap absorption, only 8.8\% of the black body radiation can be absorbed by a relaxed \GeSn{0.83}{0.17} having bandgap energy of $\sim\SI{0.29}{\electronvolt}$. This fraction increases to $\sim 86.7$\% at $\SI{2000}{\kelvin}$.}
    \label{fig:WasteHeatTemperature}
\end{figure*}

Narrow bandgap \ac{TPV} cells have been the subject of extensive investigations motivated by their strategic importance in harvesting infrared radiations for heat waste conversion, portable devices, power beaming, and space applications \cite{luInAsThermophotovoltaicCells2018,chanPortableMesoscaleThermophotovoltaic2018,wangRadioisotopeThermophotovoltaicGenerator2020, rizzoComparisonTerahertzMicrowave2020a, yangNarrowBandgapPhotovoltaic2022}. In general, \ac{TPV} cells are electronic devices consisting of a heat generator, a radiator to translate heat into an emission spectrum, an optical filter such as glass or quartz tubes, and a cell that converts infrared photon energy to electrical energy. Recently, tremendous efforts have been expended to develop mid-infrared \ac{TPV} cells to leverage their several technological advantages such as the potential to exploit the atmospheric transparency windows for eye-safer energy beaming and the ability to generate energy from the staggering amount of heat wasted by major industrial activities \cite{yangNarrowBandgapPhotovoltaic2022}. In fact, as illustrated in the chart displayed in \fref{fig:WasteHeatTemperature}{(a)}, vital manufacturing sectors such as iron, steel, cement, glass, and metallurgy suffer large energy footprints exacerbated by heat losses at temperatures below $\SI{2000}{\kelvin}$. Efficient absorption and conversion of black-body radiations emitted in this range require semiconductors with bandgap energy ($E_g$) smaller than $\sim\SI{0.7}{\electronvolt}$. Indeed, by simply examining the black-body radiation power density as a function of temperature, while neglecting sub-bandgap absorption, one can deduce that about 50.5\% of the $\SI{1000}{\kelvin}$ black-body radiation can be absorbed by a semiconductor having a bandgap energy of $\SI{0.3}{\electronvolt}$. This fraction increases to 75.5\% and 86.7\% at the temperatures of $\SI{1500}{\kelvin}$ and $\SI{2000}{\kelvin}$, respectively.

\par Compound semiconductors have been explored to implement both far-field and near-field sub-$\SI{2000}{\kelvin}$ \ac{TPV} cells reporting a measured efficiency in the 1-11\% range, while theory hints to higher values reaching 24\% \cite{luInAsThermophotovoltaicCells2018, gamelReviewThermophotovoltaicCell2021,yangNarrowBandgapPhotovoltaic2022,nftpv2022, Mialhe1900}. For instance, In$_{0.20}$Ga$_{0.80}$As$_{0.18}$Sb$_{0.82}$ ($E_g = \SI{0.5}{\electronvolt}$) \ac{TPV} cells on lattice-matched GaSb deliver an open circuit voltage (\Voc{}) of $\sim\SI{0.3}{\volt}$, a short circuit current density (\Jsc{}) up to $\SI{3}{\ampere\per\square\centi\meter}$, and an \ac{IQE} of $\sim 90\%$ \cite{maukGaSbrelatedMaterialsTPV2003,wangHighquantumefficiencyEVGaInAsSb1999,luInAsThermophotovoltaicCells2018}. GaSb \ac{TPV} cells have the potential to provide a \ac{PCE} of $\sim 16\%$ and are suitable for heat sources ranging from 1300 to $\SI{1500}{\kelvin}$. These cells exhibit an \ac{IQE} above 90\% at wavelengths up to $\SI{1.8}{\micro\meter}$ \cite{sulimaFabricationSimulationGaSb2001a, gamelReviewThermophotovoltaicCell2021}. However, the \ce{Zn} diffusion needed to form the $p$-type GaSb emitter hinders the cell performance \cite{schleglTPVModulesBased2004}. For InGaAs ($E_g = \SI{0.74}{\electronvolt}$) \ac{TPV} cells, a \ac{PCE} of 9.2-16.4\% was reported for black-body temperature in the 873-$\SI{1323}{\kelvin}$ range, but the change in the illumination spectrum was shown to have an important impact on the current conduction mechanisms that dominate the minority charge transport in these cells \cite{tanInvestigationInGaAsThermophotovoltaic2014, gamelPerformanceGe532021}. Narrower bandgap semiconductors such as InAs ($E_g = \SI{0.35}{\electronvolt}$) yield higher efficiency \ac{TPV} cells for lower temperature heat sources ($<\SI{1250}{\kelvin}$) as more radiated low-energy photons can be absorbed by these semiconductors \cite{krierLowBandgapMidinfrared2015, luInAsThermophotovoltaicCells2018}. In this regard, \ac{TPV} cells employing an InAs \pin{} diode enable a 3.6\% \ac{PCE} at a black-body temperature of $\SI{950}{\celsius}$ \cite{luInAsThermophotovoltaicCells2018}. 

\par Notwithstanding the progress achieved so far in developing narrow bandgap \ac{TPV} cells, the substrates needed to grow these cells are costly and suffer large dark currents. To mitigate the latter, narrow bandgap \ac{ICPV} based on InAs/GaSb multiple quantum wells have been recently proposed \cite{yangInterbandCascadePhotovoltaic2010, hinkeyInterbandCascadePhotovoltaic2013}. These devices alleviate the limiting factors of bulk narrow bandgap \ac{TPV} (high saturation dark current density, relatively low absorption coefficient, short diffusion length, material quality, etc.) thanks to the type II broken-gap alignment at the InAs/GaSb interface, which mainly enables carrier collection through tunneling. Under standard black-body radiation illumination \cite{yangInterbandCascadePhotovoltaic2010, hinkeyInterbandCascadePhotovoltaic2013}, the \ac{ICPV} devices operating at $\SI{80}{\kelvin}$ exhibit relatively high \comments{open-circuit voltage}\Voc{} ($\sim\SI{1.1}{\volt} > E_g/e$) demonstrating that the multiple stage absorbers operate in series. However, at 300 K the \ac{ICPV} devices perform poorly with \Voc{} around $\SI{5.7}{\milli\volt}$ and an output power density below $\SI{0.002}{\milli\watt\per\square\centi\meter}$. The deterioration is attributed to the saturation current density being significantly higher than the photo-current density generated by the radiation from the black-body source.

A cost-effective and scalable alternative to compound semiconductor mid-infrared \ac{TPV} materials would be the silicon-compatible narrow bandgap \GeSn{1-y}{y} semiconductors \cite{moutanabbirMonolithicInfraredSilicon2021}. These group IV alloys are grown on large-diameter silicon wafers and can cover the entire mid-infrared range by increasing the Sn content. Indeed, the recent progress in their epitaxial growth led to the demonstration of a variety of monolithic mid-infrared emitters and detectors \cite{atallaAllGroupIV2021b, atallaHighBandwidthExtendedSWIRGeSn2022a, bucaRoomTemperatureLasing2022, changMidinfraredResonantLight2022, chretienGeSnLasersCovering2019, chretienRoomTemperatureOptically2022, elbazUltralowthresholdContinuouswavePulsed2020b, joo1DPhotonicCrystal2021, jungOpticallyPumpedLowthreshold2022, li30GHzGeSn2021, liuSnContentGradient2022, luoExtendedSWIRPhotodetectionAllGroup2022b, marzbanStrainEngineeredElectrically2022, talamassimolaCMOSCompatibleBiasTunableDualBand2021, tranSiBasedGeSnPhotodetectors2019, xuHighspeedPhotoDetection2019, zhouElectricallyInjectedGeSn2020a}. Building on these achievements, herein we introduce \GeSn{1-y}{y} alloys  to design silicon-integrated mid-infrared \ac{TPV} cells and discuss their performance and its evolution as a function of the basic material properties.

% load the body of the text
%auto-ignore
\section{Device structure and theoretical Framework }\label{sec:modelTheory}
The proposed all-group IV mid-infrared TPV cells consist of a fully relaxed \GeSn{1-y}{y} \pin{} homojunction grown on a silicon wafer using a germanium interlayer - commonly known as a virtual substrate. \fref{fig:WasteHeatTemperature}{(b)} exhibits a schematic illustration of the device structure. Varying the Sn content between $5$ and $20\text{ at.}\%$ yields bandgap energies covering the mid-infrared range up to $\SI{6}{\micro\meter}$, as displayed in \fref{fig:WasteHeatTemperature}{(c)}. This composition range is consistent with the recent experimental studies on the growth of \GeSn{1-y}{y} alloys \cite{moutanabbirMonolithicInfraredSilicon2021}. Note that the front electrode shadowing and the parasitic resistances are neglected in evaluating the device performance. Moreover, the one-dimensional treatment is justified since the device dimensions are significantly larger than the device thickness. Thus, the performance and the electrical properties of the \ac{TPV} cell are estimated by solving self-consistently the steady-state coupled Poisson drift-diffusion equations \eqref{eqn:PoisDriftDiffEqn}: 
% \begin{align}
%     \begin{aligned}
%         \pdv{x}\left(\varepsilon_r\pdv{\phi}{x}\right) &= -\frac{e}{\varepsilon_0}\left(p + n + N_D^+ - N_A^-\right) \\
%         \pdv{x}\left(\mu_n n\pdv{E_{f_n}}{x}\right) &= e\left(R - G\right) \\
%         \pdv{x}\left(\mu_p p\pdv{E_{f_p}}{x}\right) &= -e\left(R - G\right)
%     \end{aligned}
%     \label{eqn:PoisDriftDiffEqn}
% \end{align}
\begin{align}
    \begin{aligned}
        \varepsilon_0\nabla_x\big(\varepsilon_r\nabla_x{\phi}\big) &= -e\left(p + n + N_D^+ - N_A^-\right) \\
        \nabla_x\big(\mu_n n\nabla_x E_{f_n}\big) &= e\left(R - G\right) \\
        \nabla_x\big(\mu_p p\nabla_x E_{f_p}\big) &= -e\left(R - G\right)
    \end{aligned}
    \label{eqn:PoisDriftDiffEqn}
\end{align}
where $\varepsilon_r$ is the relative dielectric permittivity, $\phi$ the electrostatic potential, and $p + n + N_D^+ - N_A^-$ is the space charge density. Besides $G - R$ is the net difference between generation and recombination while $E_{f_n}$ and $E_{f_p}$ are the electron and holes quasi-Fermi levels, respectively.
This set of equations mainly describes the carrier dynamics and transport, which requires a rigorous description of both the electrical and optical properties of the materials involved. 
In the simulations of the optical behavior and related calculations of the generation rate $G(x)$ and the absorbance $\mathcal{A}$, the complete stack of layers is considered to describe both the frontside and backside illumination. Once these properties are obtained, as a first approximation, the simulations were restricted to the set of active layers, namely the \pin{} junction (\fref{fig:WasteHeatTemperature}{(b)}).

\subsection{Optical characteristics of the TPV \texorpdfstring{\pin{}}{p-i-n} device}
\begin{figure}[htb]
% \begin{figure}[H]
    \centering
    \includegraphics[width=\columnwidth]{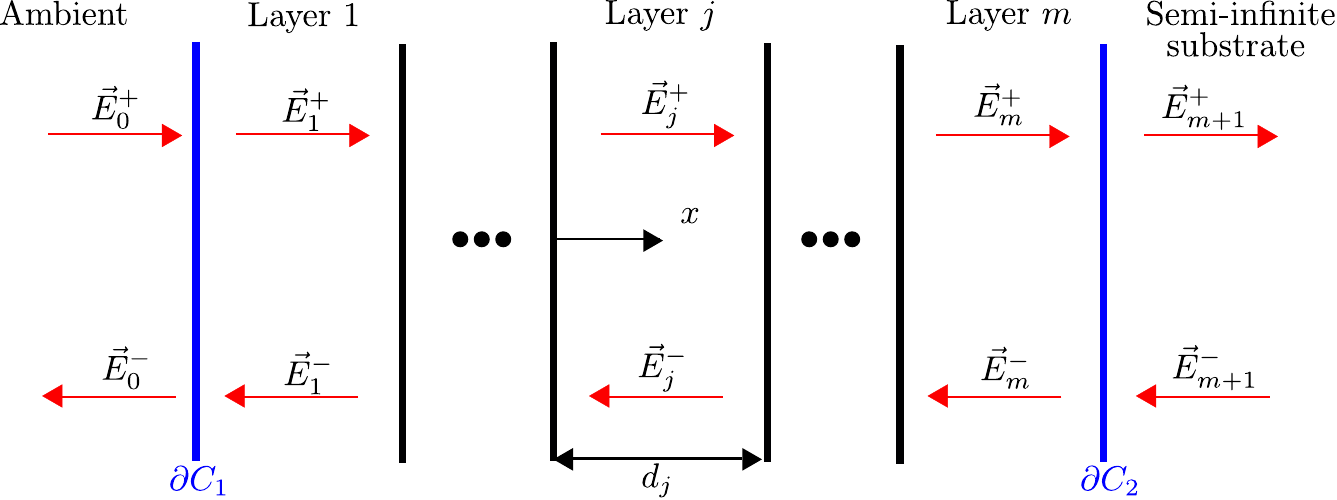}
    \caption{Diagram of a multilayered structure having $m$ layers between a semi-infinite transparent ambient and a semi-infinite substrate. The optical electric field in layer $j$ is divided into the incident wave $E_j^+$ and the reflected wave $E_j^-$. Each layer $j$ is described by its thickness $d_j$ and the complex refractive index $\tilde{\eta}_j = n_j + ik_j$.}
    \label{fig:schemMultilayer}
\end{figure}

In a single semiconductor slab, the generation rate $G(x)$ of \acpl{EHP} at a distance $x$ from the surface is usually derived from the Beer-Lambert law \cite{Nelson2003,szePhysicsSemiconductorDevices2007}. However, in a multilayered structure with different refractive indices from one layer to the others, light absorption can be significantly affected by the optical interference of the incident and the reflected waves. In this study, the \ac{OTMM} was employed to simulate light propagation through the different layers of the Ge$_{1-y}$Sn$_{y}$ \ac{TPV} cells. For a light incident from the left of the multilayered structure, as shown in \fref{fig:schemMultilayer}{}, the total electric field in a layer $j$ at an arbitrary distance $x$ to the left interface ($0\leq x\leq d_j$, where $d_j$ is the layer thickness) is given by \cite{petterssonModelingPhotocurrentAction1999}:
\begin{equation}
    E_j(x) = E_j^+(x) + E_j^-(x)
    \label{eqn:totalField}
\end{equation}

The time average of the energy dissipated per second $Q_j(x, \lambda)$ at a position $x$ in layer $j$ and at a specific wavelength $\lambda$ is given by \cite{petterssonModelingPhotocurrentAction1999}
% \begin{equation}
%     Q_j(x, \lambda) = \frac{1}{2}c\varepsilon_0\alpha_j(\lambda)n_j(\lambda)\abs{E_j(x)}^2
%     \label{eqn:dissipationCoeff}
% \end{equation}
\begin{equation}
    \begin{split}
        Q_j(x, \lambda) & = \frac{1}{2}c\varepsilon_0\alpha_j(\lambda)n_j(\lambda)\abs{E_j(x)}^2 \\
         & = \left(\frac{\alpha_jn_j}{n_0}\right)\abs{t_j^+}^2I_0\biggl[e^{-\alpha_jx} + \rho_j^{''}\cdot e^{\alpha_j(2d_j - x)} {}\\
         & + 2\rho_j^{''}\cdot e^{-\alpha_jd_j}\cdot \cos\left(\frac{4\pi n_j}{\lambda}(d_j - x) + \delta_j^{''}\right)\biggr]
         % & \quad \times\right] 
    \end{split}
     \label{eqn:dissipationCoeffFinal}
\end{equation}
where $c$ is the speed of light in free space, $\varepsilon_0$ the vacuum permittivity, $n_j$ the refractive index of layer $j$, $n_0$ the refractive index of the ambient environment, $\alpha_j = 4\pi k_j/\lambda$ the absorption coefficient of layer $j$, $k_j$ its extinction coefficient, $I_0 = (hc/\lambda)\phi_f\left(\frac{hc}{\lambda}, T_{bb}\right)$ the intensity of the incident light, and $\phi_f$ the photon flux of the incident radiation at temperature $T_{bb}$. $\phi_f$ can either be described by the Planck's law for pure black-body radiation or taken as an input parameter for more realistic gray-body emitters or even selective emitters. Moreover, $t_j^+$ is the internal transfer coefficient that relates the incident plane wave to the internal electric field propagating toward the positive $x$-direction in layer $j$, while $\rho_j^{''}$ and $\delta_j^{''}$ are the magnitude and argument of the internal complex reflection coefficient in layer $j$, respectively \cite{petterssonModelingPhotocurrentAction1999}. \comments{In equation \eqref{eqn:dissipationCoeffFinal}} \comments{For simplicity, the dependence on the wavelength $\lambda$ of $n_j$, $n_0$, and $\alpha_j$ is neglected.} 

\indent Using equation \eqref{eqn:dissipationCoeffFinal}, the generation rate $G_j(x)$ at an arbitrary point $x$ in layer $j$ is derived: 
\begin{equation}
    G_j(x) = \int_{E}\frac{Q_j(x, E)}{E}dE,\quad E = \frac{hc}{\lambda},\quad 0\leq x\leq d_j 
    \label{eqn:genRate}
\end{equation}
\indent The coherence length $L_c$ of a black-body radiation of temperature $T_{bb}$ is given as \cite{AxelDonges1998}:
\begin{equation}
    L_c = \frac{hc}{4kT_{bb}}
\end{equation}
where $k$ is the Boltzmann's constant. For the temperature range considered here (\fref{fig:WasteHeatTemperature}{(c)}), $L_c$ is estimated to be significantly smaller compared to the thickness of the silicon substrate ($\sim \SI{500}{\micro\meter}$). For this reason, the use of ordinary \ac{OTMM} will lead to narrow Fabry-Perot oscillations in the calculated reflectance, transmittance, and absorbance spectra. Thus, to account for the incoherency of light in the structure, the absorbance spectra $\mathcal{A}(\lambda)$ was computed using the \ac{GTMM}, which is valid for mixed coherent/incoherent layer stacking \cite{Katsidis02}. Moreover, the model presented here considers the high injection regime as it is relevant to broader applications including concentrator TPV modules and laser power beaming.

\begin{table*}[htb]
\setlength{\tabcolsep}{5pt}
\renewcommand{\arraystretch}{1.7}%
\label{tab:matParams}
\centering
\caption{The physical parameters used in the simulation for the \GeSn{0.83}{0.17} \pin{} \ac{TPV} cells}
\adjustbox{max width=\textwidth}{%
\begin{tabular}{ccccccccccccccccc}
\toprule
\multicolumn{1}{c}{} & \multicolumn{1}{c}{\thead{\textbf{Thickness}\\ ($\SI{}{\nano\meter}$)}} & \multicolumn{3}{c}{\thead{\textbf{Energy}\\($\SI{}{\electronvolt}$)}} & \multicolumn{2}{c}{\thead{\textbf{Effective} \\ \textbf{masses ($m_0$)} \\\cite{Assali2021}}} & \multicolumn{1}{c}{\thead{\textbf{Dielectric}\\\textbf{permittivity}\\\cite{Chang2010}}} & \multicolumn{2}{c}{\thead{\textbf{Mobility}\\ ($\SI{}{\square\centi\meter\per\volt\per\second}$)\\\cite{bosi2010,caughey1967}}} & \multicolumn{2}{c}{\thead{\textbf{Doping}\\ (cm$^{-3}$)}} & \multicolumn{1}{c}{\thead{\textbf{SRH}\\ \textbf{lifetime} ($\SI{}{\micro\second}$)}} & \multicolumn{2}{c}{\thead{\textbf{Auger}\\ \textbf{coefficients}\\($10^{-32}$ cm$^6$/s)}} & \multicolumn{2}{c}{\thead{\textbf{Surface}\\ \textbf{recombination}\\ \textbf{velocity}\\(cm/s) \cite{virgilio2013}}} \\
\cmidrule(rl){2-2} \cmidrule(rl){3-5} \cmidrule(rl){6-7} \cmidrule(rl){8-8}\cmidrule(rl){9-10}\cmidrule(rl){11-12}\cmidrule(rl){13-13}\cmidrule(rl){14-15}\cmidrule(rl){16-17}
\textbf{Material} & $d$ & $E_g$  \cite{Assali2021} & $E_p$ \cite{Chang2010} & $\chi_s$ \cite{GeAffinityIoffe}& $m_c$  & $m_{hh} (m_{lh})$ & $\varepsilon_r$ & $\mu_n$ & $\mu_p$ & $N_D$ & $N_A$ & $\tau_n = \tau_p$ & $C_n$ & $C_p$ & $S_n$ & $S_p$  \\
\midrule
$p$-layer & 500 & \multirow{3}{*}{0.291}  & \multirow{3}{*}{25.91} & \multirow{3}{*}{4} & \multirow{3}{*}{0.0243} & \multirow{3}{*}{\makecell{0.0320\\ (0.0741)}} & \multirow{3}{*}{17.36} & 3895 & 421.43 & 0 & $10^{18}$ & \multirow{3}{*}{2} & \multirow{3}{*}{3} & \multirow{3}{*}{7} & $10^{3}$ & $+\infty$\\
$i$-layer & 2000 & & & & & & & 3895 & 2505 & 0 & 0 & & & & \textcolor{blue}{*} & \textcolor{blue}{*}\\
$n$-layer & 500 & & & & & & & 1888.2 & 2505 & $10^{17}$ & 0 & & \multicolumn{2}{c}{\cite{conradt1972,marchetti2001}} & $+\infty$ & $10^{3}$ \\
\bottomrule
\end{tabular}
}
{\raggedright Note: \quot{\textcolor{blue}{*}} indicates that the parameter is not required. \par}
\end{table*}
\subsection{Numerical solution of the steady-state coupled Poisson drift-diffusion equations}
The free carrier densities $(n,\,p)$, involved in the system of equations \eqref{eqn:PoisDriftDiffEqn}, are estimated using the \ac{PBA} and the Fermi-integral of index 1/2 ($\mathcal{F}_{1/2}$) \cite{chuang2012physics}. Moreover, the resolution of the Poisson equation using the Newton-Raphson method, involves the compution of the derivative of $\mathcal{F}_{1/2}$ which is proportional to $\mathcal{F}_{-1/2}$. The numerical computation of the Fermi-integrals in a multilayered structure is a time-consuming and memory-intensive process. Here, we use an approximation of $\mathcal{F}_j$ to simplify the calculation process \cite{Humet1983}:
\begin{equation}
    \label{eqn:FermiIntegralApprox}
    \mathcal{F}_{j-1}(y)=\left(\frac{j\cdot 2^{j}}{\left[b+y+\left(|y-b|^{c}+a^{c}\right)^{1 / c}\right]^{j}}+\frac{e^{-y}}{\Gamma(j)}\right)^{-1}
\end{equation}
where $a = \left(1 + 3.75j+0.025j^2\right)^{1/2}$, $b = 1.19+0.61j$,  $c = 2\left(1 +(2-\sqrt{2}) 2^{-j}\right)$ and $\Gamma$ is the Gamma function. More recent approaches have been presented in the literature for the computation of the Fermi integrals \cite{Guseinov_2010,fukushima2014417}. However, this approximation has the merit to allow a simple calculation of the $\mathcal{F}_{j}(y)$ for any real value of $j$ and any $y$ using only one expression and with acceptable accuracy.

\par A unique solution of Poisson's equation requires the identification of the boundary conditions at the metal-semiconductor interfaces. Herein, different contact models are considered. An ideal ohmic contact is specified by assuming a vanishing electric field at the boundaries of the structure, and therefore, zero von Neumann boundary conditions.\comments{ are applied.In that case, von Neumann boundary conditions are applied for the electrostatic potential $\phi$ as presented in equation \eqref{eqn:idealOhm}\cite{andlauer2009}.}
Schottky contacts are also established by employing Dirichlet boundary conditions for the electrostatic potential $\phi$. The latter depends on the work functions $\phi_m$ of the metals, the voltage bias, the electron affinity $\chi_s$ of the semiconductors, and the thermal equilibrium Fermi level.

\par In Si solar cells, the impact of the intrinsic recombination mechanisms (Auger and radiative recombination) on the conversion efficiency was shown to be negligible \cite{Richter_2013}. However for narow-bandgap \GeSn{1-y}{y} materials, depending on the material quality, these processes could become the most dominant. For that reason, the established formalism takes into account Auger, \ac{SRH}, and radiative recombination mechanisms.
% \par Regarding the recombination processes, the established formalism takes into account Auger, \ac{SRH}, and radiative recombination mechanisms. 
The net non-radiative recombination rates are computed following the process described in \cite{chuang2012physics}. As for the radiative recombination, rather than using the bimolecular recombination coefficient $B$ to estimate the net rate, equation \eqref{eqn:radRate} below is used, where  $E_{f_n}$ and $E_{f_p}$ are the quasi-Fermi levels, $P$ is the Kane's parameter, $e$ is the elementary charge, $n_r$ is the refractive index of the material, $E_v^0 = E_v - e\phi - (\hbar^2/2m_v)k_{0,v}$, $E_c^0 = E_c - e\phi + (\hbar^2/2m_c)k_{0,v}$, with $\phi$ the electrostatic potential, and $f$ the Fermi-Dirac distribution \cite{chuang2012physics}. Besides, $k_{0,v} = (2m^*_{c,v}/\hbar^2)^{1/2}\sqrt{\hbar\omega - E_c + E_v}$, with $m^*_{c,v}$ the reduced effective mass. 

\begin{equation}
    \begin{aligned}
    r_{\mathrm{sp}}(\hbar \omega)=& \left(\frac{P^{2}n_{r} e^{2} \hbar \omega}{6\pi^3 \hbar^{4} c^{3} \varepsilon_{0}}\right) \sum_{v}\left(\frac{2 m^*_{c,v}}{\hbar^{2}}\right)k_{0,v} \\
    & \times f\left(\frac{E_{c}^{0}-E_{f_{n}}}{k_{B} T}\right) f\left(\frac{E_{f_{p}}-E_{v}^{0}}{k_{B} T}\right)\\
    R_{\mathrm{rad}}(x) =& \int_{0}^{+\infty}\biggl[r_{\mathrm{sp}}(E, x) - r_{\mathrm{sp}}^{\text{eq}}(E, x)\biggr] \mathrm{d} E
    \end{aligned}
    \label{eqn:radRate}
\end{equation}
This equation should be more general since the bimolecular recombination approximation is only accurate for non-degenerate materials where the quasi-Fermi levels are very close to the middle of the bandgap \cite{RadLifetimeArxiv2023}. 

\par The drift-diffusion equation is solved numerically under different boundary conditions imposed on the quasi-Fermi levels at both electrical contacts. Indeed, at these boundaries, the values of the quasi-Fermi levels are mainly defined by the current density $j_{n,p}$ and the surface recombination velocities $S_{n,p}$ for electrons and holes at the contacts, as specified by equation \eqref{eqn:boundCondDD}, in which $n_{eq}$ and $p_{eq}$ are the carrier densities at thermal equilibrium \cite{Nelson2003}. 
\begin{align}
    \begin{aligned}
        j_n(\partial C_1) &= eS_{n, 1}\Big(n(\partial C_1) - n_{eq}(\partial C_1)\Big)\\
        j_n(\partial C_2) &= -eS_{n, 2}\Big(n(\partial C_2) - n_{eq}(\partial C_2)\Big)\\
        j_p(\partial C_1) &= -eS_{p, 1}\Big(p(\partial C_1) - p_{eq}(\partial C_1)\Big)\\
        j_p(\partial C_2) &= eS_{p, 2}\Big(p(\partial C_2) - p_{eq}(\partial C_2)\Big)
    \end{aligned}
    \label{eqn:boundCondDD}
\end{align}
% \begin{align}
%     \begin{aligned}
%         j_n\Bigr\rvert_{\partial C_1} &= eS_{n, 1}\left(n\Bigr\rvert_{\partial C_1} - n_{eq}\Bigr\rvert_{\partial C_1}\right)\\
%         j_n\Bigr\rvert_{\partial C_2} &= -eS_{n, 2}\left(n\Bigr\rvert_{\partial C_2} - n_{eq}\Bigr\rvert_{\partial C_2}\right)\\
%         j_p\Bigr\rvert_{\partial C_1} &= -eS_{p, 1}\left(p\Bigr\rvert_{\partial C_1} - p_{eq}\Bigr\rvert_{\partial C_1}\right)\\
%         j_p\Bigr\rvert_{\partial C_2} &= eS_{p, 2}\left(p\Bigr\rvert_{\partial C_2} - p_{eq}\Bigr\rvert_{\partial C_2}\right)
%     \end{aligned}
%     \label{eqn:boundCondDD}
% \end{align}
From these equations, the minority carriers that reach the surfaces $\partial C_1$ and $\partial C_2$ are supposed to recombine\comments{there}. However, for infinite values of surface recombination, the minority carrier density at the surface is defined by its thermal equilibrium value, and the quasi-Fermi level is therefore given by the thermal equilibrium Fermi level and the voltage bias applied. This situation is usually seen for $p$-$n$ junctions where the quasi-Fermi levels reach the same value at both ends of the structure \cite{szePhysicsSemiconductorDevices2007,streetman2000}. 
% \begin{itemize}
%     % \item \textbf{\textcolor{green!50!black}{Is the equation for this kind of boundary condition required here? Refs for the textbooks?}}
%     \item \textbf{\textcolor{green!50!black}{Refs for the textbooks?}}
% \end{itemize}
\par The system of coupled differential equations \eqref{eqn:PoisDriftDiffEqn} is solved in a block-iterative way following the procedure described in \cite{hackenbuchner2002}. This approach consists of alternating solutions of Poisson's equation with fixed quasi-Fermi levels and the current equations with fixed electrostatic potential. The equations are repeatedly solved in each block until convergence is achieved before proceeding with the following outer iteration step in the other block. The $J$-$V$ characteristics of the \ac{TPV} cell are extracted after solving \eqref{eqn:PoisDriftDiffEqn} for different values of the voltage bias $V$. Afterward, \Voc{}, \Jsc{}, the \ac{FF}, and the \ac{PCE} are computed. In this study, the \ac{PCE} is defined as the ratio of the output electrical power to the incident power. However, to compare with data reported in the literature, the ratio of the output electrical power to the power absorbed by the device is also computed. This quantity is sometimes named the absorption efficiency ($\eta_{\text{abs}}$) or the pairwise \ac{PCE}, and it is appropriate for laboratory scale, experimental materials \cite{burgerPresentEfficienciesFuture2020a}. The physical parameters relevant to device simulations are summarized in Table~\ref{tab:matParams}. 

\begin{figure}[H]
% \begin{figure}[htb]
    \centering
    \includegraphics[width=.8\columnwidth, keepaspectratio]{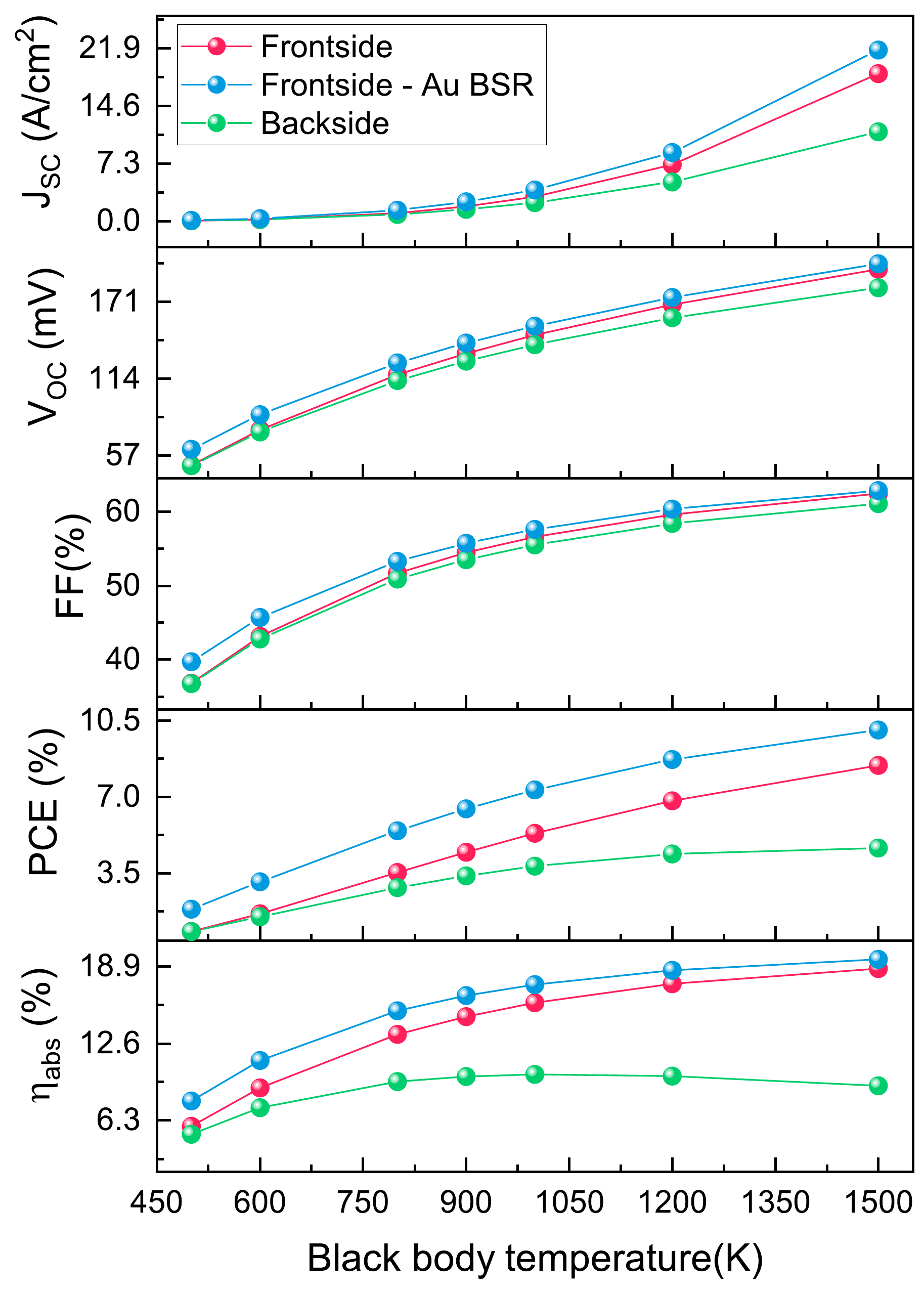}
    \caption{Impact of the black-body temperature $T_{bb}$ on the power conversion efficiency of the  \GeSn{0.83}{0.17} \pin{} \ac{TPV}.}
    \label{fig:resTbbVar_idealContact}
\end{figure}

\section{Results and discussion} \label{sec:res_dic}
The simulations are focused on \GeSn{0.83}{0.17} heterostructures corresponding to a bandgap energy of 0.291 eV. The \GeSn{0.83}{0.17}-based \ac{TPV} cells, operating at $\SI{300}{\kelvin}$, are illuminated by black-body radiation sources whose temperature $T_{bb}$ is within the $\SI{500}{\kelvin}$ to $\SI{1500}{\kelvin}$ range. 

Assuming ideal ohmic contacts, the evolution of the \ac{TPV} electrical parameters with the radiation temperature is studied, for both front-side and back-side illuminations, and the results are presented in \fref{fig:resTbbVar_idealContact}{}. 

With the material parameters provided in Table \ref{tab:matParams}, the short-circuit current \Jsc{} is mainly dominated by the absorption of the incident radiation and the generation rate. At a bandgap energy of $\sim\SI{0.291}{\electronvolt}$, only a small part of the black-body radiation emitted at 500 K is absorbed by the \ac{TPV} cell, leading to relatively low values of generation rate and, therefore, a small \Jsc{} of $\SI{72.92}{\milli\ampere\per\square\centi\meter}$ for front-side illumination. By increasing the temperature $T_{bb}$, the spectral radiant emittance of the source shifts to higher energies where more and more emitted photons are absorbed by the cell. In that situation, more \acpl{EHP} are generated and collected in the \pin{} structure, contributing to the increase of the charge current, which reaches $\SI{18.68}{\ampere\per\square\centi\meter}$ at $\SI{1500}{\kelvin}$. 
 
 % \begin{figure}[htb]
\begin{figure}[H]
    \includegraphics[width=\columnwidth]{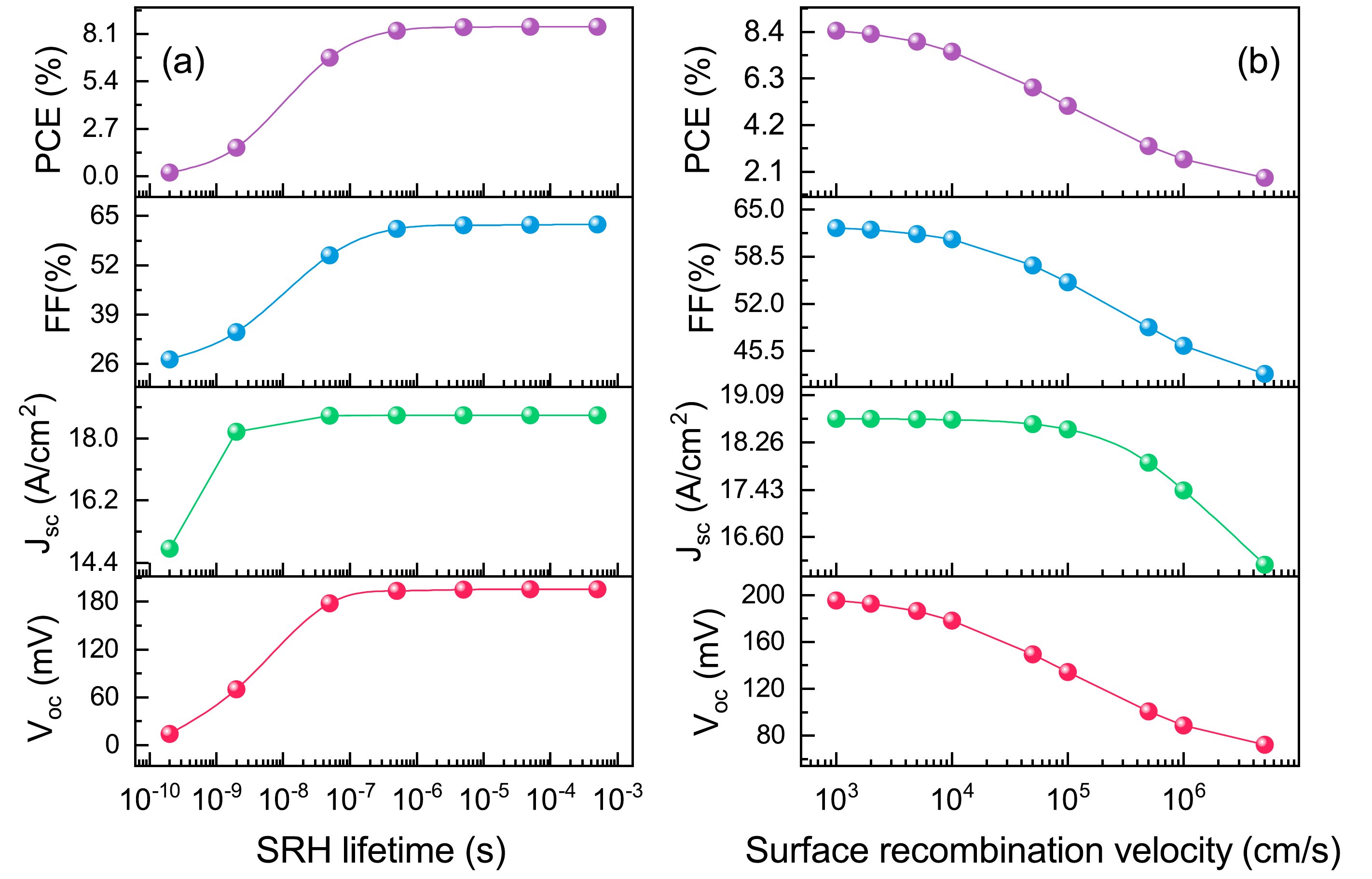}
    \caption{Impact of the \ac{SRH} lifetimes (a) and the surface trapping centers (b) on the key parameters of the \GeSn{0.83}{0.17} \pin{} \ac{TPV} at $T_{bb} = \SI{1500}{\kelvin}$.}
    \label{fig:resDefectVar}
\end{figure}

\par For an open circuit, no current flows across the device, and thus all the photo-generated charge carriers must recombine. Since the number of generated \acpl{EHP} increases with the radiation temperature, the open-circuit recombination rate must be large enough to cancel out the generation rate and the current. For this to occur, the energy offset between the quasi-Fermi levels increases therefore leading to higher values of \Voc{}, as shown in the bottom panel of \fref{fig:resTbbVar_idealContact}{}. Indeed, \Voc{} increases from $\SI{49.6}{\milli\volt}$ at $\SI{500}{\kelvin}$ to reach about $\SI{195}{\milli\volt}$ at $\SI{1500}{\kelvin}$. At the same time, the maximum output power of the \ac{TPV} cell increases with the source temperature, enabling significant improvements in the \ac{FF} and the \ac{PCE}. At $\SI{1500}{\kelvin}$, for instance, a \ac{FF} of $\sim 62.5\%$ is estimated in addition to a \ac{PCE} of about $8.5\%$. Starting from $\SI{1000}{\kelvin}$, the difference between the \ac{TPV} performance parameters under front-side and back-side illuminations starts to be noticeable. This difference can be explained by the higher number of \acpl{EHP} generated in the $i$-layer when the radiation is sent to the top of the device. With the total recombination rate $R(x)$  being negligible compared to $G(x)$ at $\SI{0}{\volt}$, more \acpl{EHP} in the absorbing layer would lead to more current and thus to higher \Jsc{}. For an open circuit, the increased number of \acpl{EHP} will also lead to higher recombination rates and higher voltage \Voc{}. The differences in these two parameters, combined with the maximum output electrical power $P_m$, induces the behavior observed for the \ac{FF}, the \ac{PCE}, and the absorption efficiency $\eta_{\text{abs}}$, even though the difference appears earlier around $\SI{600}{\kelvin}$ and is more pronounced for $\eta_{\text{abs}}$.

\par Besides, the concept of transmissive spectral control and its impact on the \ac{TPV} efficiency is also investigated. Herein, a layer of gold is added at the bottom of the \ce{Si} substrate to act as a \ac{BSR} and mainly reflect radiation with energy lower than the \GeSn{0.83}{0.17} bandgap, while absorbing radiation with higher energies \cite{wernsmanGreater20Radiant2004}. The addition of this layer induces an increase of the sub-bandgap reflectance, with the average value going from $\sim 50\%$ to $\sim 80\%$. At the same time, there is an increase of the absorption of radiation with higher energies, leading to the creation of more \acpl{EHP}. However, only a slight overall improvement is observed, as shown in \fref{fig:resTbbVar_idealContact}{}. Better improvements should be observed with the use of a more appropriate reflector (with a possible addition of a spacer) in our specific range of energy.

\begin{table*}[htbp]
    \setlength{\tabcolsep}{10pt}
    \renewcommand{\arraystretch}{1.7}%
    \centering
    \adjustbox{max width=\textwidth}{%
    \begin{threeparttable}
    \caption{Performance parameters of current narrow bandgap \ac{TPV} cells.}
    \label{tab:perf_TPV}
    \begin{tabular}{@{}ccccccccccc@{}} 
      \toprule
      \thead{Structure} & \thead{Absorber material} & \thead{$E_g$\\($\SI{}{\electronvolt}$)} & \thead{$T_c$ \\ (K)} & \thead{$P_{\text{in}}$ \\ (W/cm$^2$)} & \thead{\Jsc{} \\ (A/cm$^2$)} & \thead{\Voc{} \\ (V)} & \thead{FF \\ (\%)} & \thead{$\eta$ \\ (\%)} & \thead{$P_{\text{out}}$ \\ (mW/cm$^2$)}& \thead{Ref}\\
       \midrule
       % $p$-$n$-$n$ & InAs$_{0.84}$Sb$_{0.03}$P$_{0.13}$ & 0.35 & 294 & \makecell{1 \\ 3 \\ 5.8} & \makecell{0.15 \\ 0.49 \\ 0.94} & \makecell{0.06 \\ 0.09 \\ 0.11} & \makecell{39.15 \\ 45.44 \\ 48.3} & \makecell{0.35 \\ 0.66 \\ 0.84} & \makecell{3.51 \\ 19.76 \\ 48.71}&\cite{maukLowbandgapEVInAsSbP2003}\\
       $p$-$n$-$n$ & InAs$_{0.84}$Sb$_{0.03}$P$_{0.13}$ & 0.35 & 294 & 3 & 0.49 & 0.09 & 45.44 & 0.66 & 19.76 &\cite{maukLowbandgapEVInAsSbP2003}\\
       $p$-$n$-$n$ & GaInAsSbP & 0.35 & 300 & 0.5 & 0.29 & 0.03 & 33 & 0.53 & 2.66 & \cite{cheethamLowBandgapGaInAsSbP2011}\\
       $p$-$i$-$n$ & InAs & 0.35 & 300 & \makecell{0.32 \\ 0.72} & \makecell{0.21 \\ 0.89} & \makecell{0.02 \\ 0.06} & \makecell{28.09 \\ 37} & \makecell{0.33 \\ 3} & \makecell{1.05 \\ 21.6} & \makecell{\cite{luInAsThermophotovoltaicCells2018}\\ \cite{krierLowBandgapMidinfrared2015}}\\
       % $p$-$i$-$n$ & InAs & 0.35 & 300 & 0.32 & 0.21 & 0.018 & 28.09 & 0.33 & 1.05 & \cite{luInAsThermophotovoltaicCells2018}\\
       % Superlattices & \makecell{InAs/GaSb \\ multi-stages} & 0.24 & 80 & 0.67 & 0.046 & 1.12 & 61.68 & 4.72 & 31.65 & \cite{yangInterbandCascadePhotovoltaic2010}\\ 
       % Superlattices & \makecell{InAs/GaSb \\ multi-stages} & 0.31 & 80 & 0.0075 & \SI{3.4e-4}{} & 1.47 & 44.56 & 2.98 & 0.22 & \cite{hinkeyInterbandCascadePhotovoltaic2013}\\ 
       % Superlattices\tnoteMod{a} & \makecell{InAs/GaSb \\ multi-stages} & 0.25 & 300 & 19 & 1.401 & 0.651 & 43.17 & 2.1 & 393.51 & \cite{Lotfi2013}\\ 
       % Superlattices\tnoteMod{a} & \makecell{InAs/GaSb \\ multi-stages} & 0.39 & 300 & 188 & 43.42 & 0.799 & 51.21 & 9.45 & \SI{1.776e4}{} & \cite{Lotfi2017}\\
       % Superlattices\tnoteMod{a} & \makecell{InAs/GaSb \\ 6 stages \\ 16 stages \\ 23 stages} & 0.23 & 300 & 17 & \makecell{\\ 3.561 \\ 1.175 \\ 0.877} & \makecell{\\ 0.347 \\ 0.919 \\ 1.459} & \makecell{\\ 35.45 \\ 36.04 \\ 36.75} & \makecell{\\ 2.57 \\ 2.29 \\ 2.77} & \makecell{\\ 437.57 \\ 388.75 \\ 470.26} & \cite{huangPerformanceAnalysisNarrowbandgap2020}\\
       % $p$-$i$-$n$\tnoteMod{} & \makecell{\GeSn{0.83}{0.17}\\\SI{400}{\nano\meter}} & 0.291 & 300 & 26.9603 & 11.51 & 0.204 & 64.08 & 5.59 & \SI{1.51e3}{} & \makecell{\textcolor{blue}{This}\\\textcolor{blue}{work}}\\
       % $p$-$i$-$n$\tnoteMod{} & \makecell{\GeSn{0.83}{0.17}\\\SI{2000}{\nano\meter}} & 0.291 & 300 & 26.9603 & 18.68 & 0.195 & 62.47 & 8.45 & \SI{2.28e3}{} & \makecell{\textcolor{blue}{This}\\\textcolor{blue}{work}}\\
       $p$-$i$-$n$\tnoteMod{} & \makecell{\GeSn{0.83}{0.17}\\\SI{2000}{\nano\meter}} & 0.29 & 300 & 0.3 & 0.21 & 0.08 & 42.95 & 2.24 & 6.67 & \makecell{\textcolor{blue}{This}\\\textcolor{blue}{work}}\\
       $p$-$i$-$n$\tnoteMod{} & \makecell{\GeSn{0.83}{0.17}\\\SI{2000}{\nano\meter}} & 0.29 & 300 & 3 & 2.08 & 0.14 & 54.96 & 5.17 & 154.97 & \makecell{\textcolor{blue}{This}\\\textcolor{blue}{work}}\\
       Superlattices\tnoteMod{a} & \makecell{InAs/GaSb \\ 6 stages} & 0.22 & 300 & 17 & 3.58 & 0.35 & 35.41 & 2.59 & 440.74 & \cite{huangPerformanceAnalysisNarrowbandgap2020}\\
       % Superlattices\tnoteMod{a} & \makecell{InAs/GaSb \\ 6 stages \\ 23 stages} & \makecell{\\0.22\\0.25} & \makecell{\\300} & \makecell{\\17} & \makecell{\\ 3.58 \\ 0.87} & \makecell{\\ 0.35 \\ 1.46} & \makecell{\\ 35.41 \\ 36} & \makecell{\\ 2.59 \\ 2.70} & \makecell{\\ 440.74 \\ 458.88} & \makecell{\\\cite{huangPerformanceAnalysisNarrowbandgap2020}}\\

       \bottomrule
    \end{tabular}
    \begin{tablenotes}
        \item [a] Unlike the other studies in which broadband sources were used to illuminate the devices, selective emitters (Interband cascade lasers) were used for the devices in the  work in \cite{huangPerformanceAnalysisNarrowbandgap2020} 
       
    \end{tablenotes}
    \end{threeparttable}
    }
\end{table*}

\begin{figure}[htpb]
% \begin{figure}[H]
    \centering
    \includegraphics[width=.9\columnwidth, keepaspectratio]{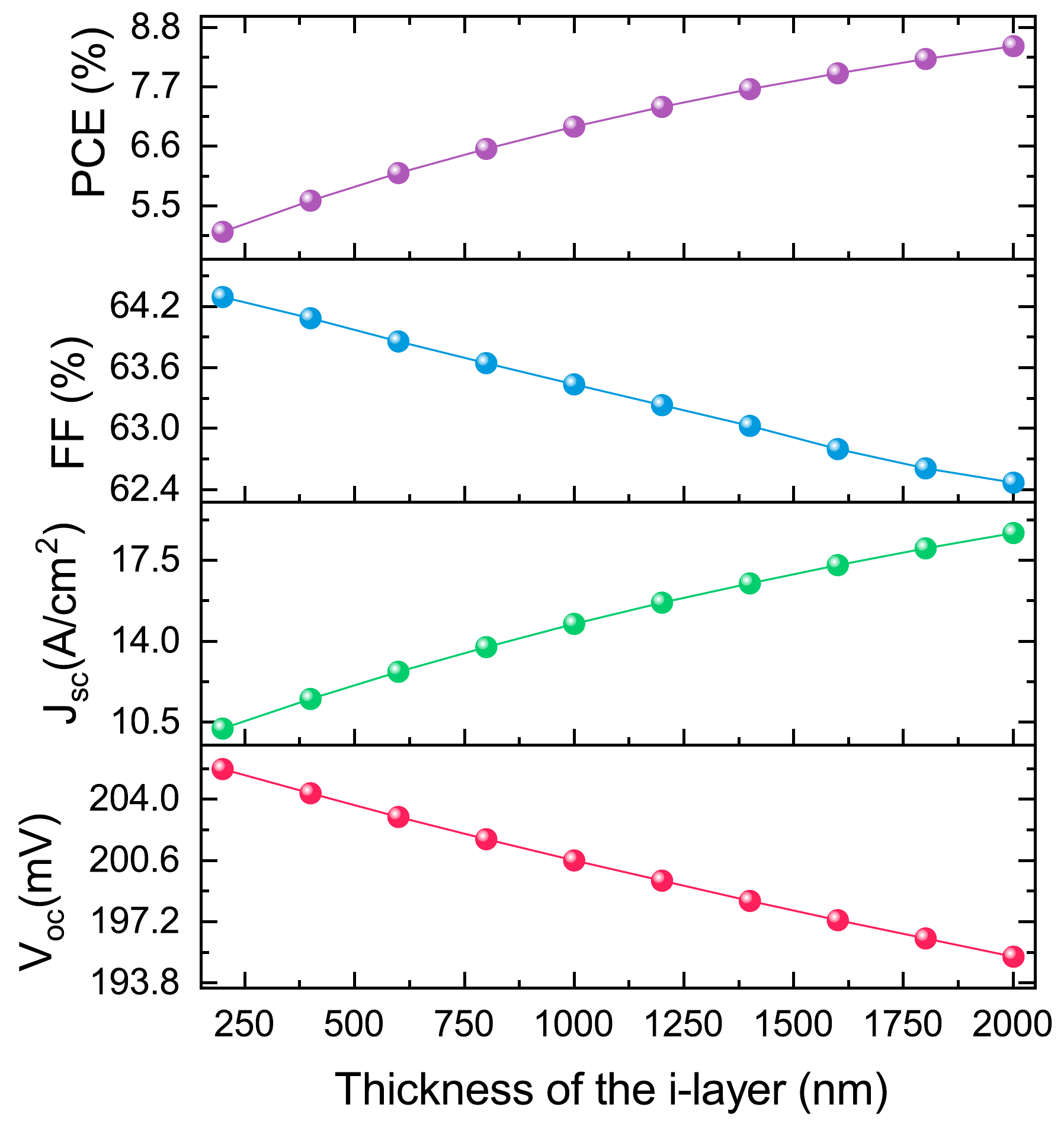}
    \caption{Impact of the thickness of the $i$-layer upon the power conversion efficiency of the \GeSn{0.83}{0.17} \pin{} \ac{TPV}. }
    \label{fig:resThicknessVar}
\end{figure}

\par The impact of different material parameters on the performance of the \GeSn{}{}-\ac{TPV} device is also investigated. Herein, devices operating at $\SI{300}{\kelvin}$ with a black-body radiation at $\SI{1500}{\kelvin}$ are considered under the front-side illumination.  At first, the \ac{SRH} lifetimes $\tau_n$ and $\tau_p$ were varied between $\SI{200}{\pico\second}$ and $\SI{1}{\milli\second}$ to reveal the potential effect of growth defects on the \ac{TPV} performance. For the sake of simplification and to reduce the number of parameters involved, $\tau_n$ was assumed to be the same as $\tau_p$, and the obtained results are displayed in \fref{fig:resDefectVar}{(a)}. 
For relatively small values of $\tau_n$ and $\tau_p$ (i.e., high defect density across the \ac{TPV} structure), most of the photo-generated carriers are trapped by the defects inside the bandgap, resulting in a degradation of the performance as seen at $\SI{200}{\pico\second}$ corresponding to a \ac{PCE} of $\sim 0.21\%$. Expectedly, better performances are obtained as lifetimes increase. However, the performance saturates at a threshold around $\SI{1}{\micro\second}$ beyond which the increase in lifetime does not yield any meaningful improvement. From this value, the radiative recombination completely takes over the \ac{SRH} recombination in the intrinsic layer, with a slight variation of the rate from one value of lifetime to the other, therefore explaining the saturation. 
% \textcolor{red}{how do you explain this saturation? (Anis)}

\par Moreover, using \ac{SRH} lifetimes of $\SI{2}{\micro\second}$, the impact of the quality of the interfaces at the metallic contacts on the \ac{TPV} performance was investigated by varying the surface recombination velocities $S_n$ of the $p$-layer and $S_p$ of the $n$-layer from $10^3$ to $10^7\SI{}{\centi\meter\per\second}$. Here, again, $S_n$ was assumed to be equal to $S_p$ to simplify the analysis. High values of surface recombination velocity result from a high number of surface trapping centers per unit area in the border region and thus correspond to poor contact quality and degradation of the \ac{TPV} performance, as shown in \fref{fig:resDefectVar}{(b)}\comments{\ref{fig:resSurfRecombVar}}. In fact, the \ac{PCE} decreases from $\sim 8.45\%$ obtained at a surface recombination velocity of $10^3\SI{}{\centi\meter\per\second}$ to reach $1.83\%$ at $S_{n,p} = \SI{5e6}{\centi\meter\per\second}$. Similar behavior is also observed for the other parameters even though \comments{the short-circuit current}\Jsc{} varies very slowly within this range of surface recombination velocities. 
This overall degradation indicates the significance of a good contact quality in achieving efficient \ac{TPV} cells. The cell efficiency can be improved by passivating the surface to reduce the number of trapping centers and, at the same time, the recombination of the photo-generated carriers at the boundary regions \cite{bose2018}.

\par The effect of the $i$-layer thickness ($t_i$) on the performance of the \ac{TPV} cell is crucial to both the overall absorption of the incident light and to the volume within which minority carriers can recombine. Using \ac{SRH} lifetimes of $\SI{2}{\micro\second}$ and surface recombination velocities of $10^3\SI{}{\centi\meter\per\second}$, the impact of the thickness of the $i$-layer on the \ac{TPV} performance was investigated. Herein, $t_i$ was varied from $200$ to $\SI{2000}{\nano\meter}$, and the obtained results are displayed in \fref{fig:resThicknessVar}{}.  The fraction of black body radiations absorbed by the \ac{TPV} cell decreases when the thickness of the $i$-layer is reduced. In this case, fewer \acpl{EHP} are generated, and \comments{the short-circuit current}\Jsc{} is reduced, resulting in lower power density and lower \ac{PCE}. The relative change is very small in the \ac{FF} ($\sim 3\%$ change) and $V_{\text{OC}}$ (5\%).
% \textcolor{red}{I am still thinking about the interpretation of the behavior of \Voc{}.}

\par The investigated \GeSn{}{}-based \ac{TPV} cells are predicted to perform relatively well in comparison to compound semiconductors. For instance, under an incident power density $P_{\text{in}}$ of $\SI{3}{\watt\per\square\centi\meter}$ and at an emitter temperature of $\SI{1500}{\kelvin}$,  a \ac{PCE} of $5.17\%$ and an output power density $P_{\text{out}}$ of $\SI{154.97}{\milli\watt\per\square\centi\meter}$ are observed for \GeSn{}{}-based devices. Under identical conditions, InAsSbP-based \ac{TPV} cells deliver an efficiency of $0.66\%$ and $P_{\text{out}}$ of $\SI{19.76}{\milli\watt\per\square\centi\meter}$, as displayed in \fref{tab:perf_TPV}{}. 
Similar behavior is also observed at lower values of $P_{\text{in}}$, where the \GeSn{}{}-based device is shown to outperform both InAs- and GaInAsSbP-based devices (\fref{tab:perf_TPV}{}). In addition to conventional \pin{} or \pnn{} junction structures, the performance of narrow bandgap \ac{ICPV} devices is also presented in \fref{tab:perf_TPV}{}. These devices benefit from the type-II broken gap alignment at the InAs/GaSb interface, which mainly enables tunneling between different cascade stages in multi-stage devices. To mimic the characteristics of a selective emitter with a narrow
emission spectrum, useful for reducing the thermalization and below-bandgap losses in a \ac{TPV} system, a type-II \ac{ICL} with emission wavelengths near $\SI{4.3}{\micro\meter}$ was employed to illuminate the different devices. In addition to reducing the losses, the \ac{ICL} offers much higher incident power density compared to black body sources \cite{hinkeyInterbandCascadePhotovoltaic2013}, leading therefore to more efficient \ac{ICPV} with relatively high output power.

% load the conclusion
% \input{3_conclusion.tex}
\section{CONCLUSION}\label{sec:concl}
This work presents an all-group IV mid-infrared thermophotovoltaic cell consisting of a fully relaxed GeSn double heterostructure with a bandgap energy of $\SI{0.29}{\electronvolt}$. A new theoretical framework was introduced to evaluate the cell performance including an accurate description of the black-body radiation absorption and of the intrinsic recombination mechanisms. This study revealed that the \GeSn{}{} cells can deliver efficiencies of up to 9\%. A slight improvement is observed under the front side illumination but vanishes below $\SI{800}{\kelvin}$. The use of a backside reflector was found to enhance cell efficiency. The impact of the material properties on the cell performance was also investigated in terms of the effects of thickness, \ac{SRH} lifetime, and surface recombination velocity. Although reducing the layer thickness from $\SI{2000}{\nano\meter}$ to $\SI{250}{\nano\meter}$ yields a relative decrease of less than 40\% in the conversion efficiency, the latter is significantly more sensitive to \ac{SRH} lifetimes and surface recombination velocity. In fact, a low lifetime has a detrimental impact on the heterostructure, thus hinting at the importance of achieving high material quality to exploit the full potential of \GeSn{}{} cells. These silicon-integrated devices can be a cost-effective alternative to current narrow bandgap compound semiconductor technologies.

% load the acknowledgments
% \bigskip \medskip
\medskip
\paperSection{ACKNOWLEDGEMENTS}
O.M.~acknowledges support from NSERC Canada (Discovery, SPG, and CRD Grants), Canada Research Chairs, Canada Foundation for Innovation, Mitacs, PRIMA Qu\'ebec, Defense Canada (Innovation for Defense Excellence and Security, IDEaS), the European Union's Horizon Europe research and innovation programme under grant agreement No 101070700 (MIRAQLS), and the US Army Research Office Grant No. W911NF-22-1-0277. R.S.~acknowledges support from the United States AFOSR on grant FA9550-21-1-0347.\\

\paperSection{AUTHORS INFORMATION}
Corresponding Authors:\\
\textcolor{blue}{$^{*}$} \href{mailto:oussama.moutanabbir@polymtl.ca}{oussama.moutanabbir@polymtl.ca}\\
Notes:\\
The authors declare no competing financial interest.

\bigskip

\bibliography{main.bbl} % Tell bibtex which .bib file to use (this one is some example file in TexLive's file tree)

\begin{thebibliography}{65}%
\makeatletter
\providecommand \@ifxundefined [1]{%
 \@ifx{#1\undefined}
}%
\providecommand \@ifnum [1]{%
 \ifnum #1\expandafter \@firstoftwo
 \else \expandafter \@secondoftwo
 \fi
}%
\providecommand \@ifx [1]{%
 \ifx #1\expandafter \@firstoftwo
 \else \expandafter \@secondoftwo
 \fi
}%
\providecommand \natexlab [1]{#1}%
\providecommand \enquote  [1]{``#1''}%
\providecommand \bibnamefont  [1]{#1}%
\providecommand \bibfnamefont [1]{#1}%
\providecommand \citenamefont [1]{#1}%
\providecommand \href@noop [0]{\@secondoftwo}%
\providecommand \href [0]{\begingroup \@sanitize@url \@href}%
\providecommand \@href[1]{\@@startlink{#1}\@@href}%
\providecommand \@@href[1]{\endgroup#1\@@endlink}%
\providecommand \@sanitize@url [0]{\catcode `\\12\catcode `\$12\catcode
  `\&12\catcode `\#12\catcode `\^12\catcode `\_12\catcode `\%12\relax}%
\providecommand \@@startlink[1]{}%
\providecommand \@@endlink[0]{}%
\providecommand \url  [0]{\begingroup\@sanitize@url \@url }%
\providecommand \@url [1]{\endgroup\@href {#1}{\urlprefix }}%
\providecommand \urlprefix  [0]{URL }%
\providecommand \Eprint [0]{\href }%
\providecommand \doibase [0]{https://doi.org/}%
\providecommand \selectlanguage [0]{\@gobble}%
\providecommand \bibinfo  [0]{\@secondoftwo}%
\providecommand \bibfield  [0]{\@secondoftwo}%
\providecommand \translation [1]{[#1]}%
\providecommand \BibitemOpen [0]{}%
\providecommand \bibitemStop [0]{}%
\providecommand \bibitemNoStop [0]{.\EOS\space}%
\providecommand \EOS [0]{\spacefactor3000\relax}%
\providecommand \BibitemShut  [1]{\csname bibitem#1\endcsname}%
\let\auto@bib@innerbib\@empty
%</preamble>
\bibitem [{\citenamefont {Lu}\ \emph {et~al.}(2018)\citenamefont {Lu},
  \citenamefont {Zhou}, \citenamefont {Krysa}, \citenamefont {Marshall},
  \citenamefont {Carrington}, \citenamefont {Tan},\ and\ \citenamefont
  {Krier}}]{luInAsThermophotovoltaicCells2018}%
  \BibitemOpen
  \bibfield  {author} {\bibinfo {author} {\bibfnamefont {Q.}~\bibnamefont
  {Lu}}, \bibinfo {author} {\bibfnamefont {X.}~\bibnamefont {Zhou}}, \bibinfo
  {author} {\bibfnamefont {A.}~\bibnamefont {Krysa}}, \bibinfo {author}
  {\bibfnamefont {A.}~\bibnamefont {Marshall}}, \bibinfo {author}
  {\bibfnamefont {P.}~\bibnamefont {Carrington}}, \bibinfo {author}
  {\bibfnamefont {C.-H.}\ \bibnamefont {Tan}},\ and\ \bibinfo {author}
  {\bibfnamefont {A.}~\bibnamefont {Krier}},\ }\href
  {https://doi.org/10.1016/j.solmat.2017.12.031} {\bibfield  {journal}
  {\bibinfo  {journal} {Solar Energy Materials and Solar Cells}\ }\textbf
  {\bibinfo {volume} {179}},\ \bibinfo {pages} {334} (\bibinfo {year}
  {2018})}\BibitemShut {NoStop}%
\bibitem [{\citenamefont {Chan}\ \emph {et~al.}(2018)\citenamefont {Chan},
  \citenamefont {Stelmakh}, \citenamefont {Karnani}, \citenamefont {Waits},
  \citenamefont {Soljacic}, \citenamefont {Joannopoulos},\ and\ \citenamefont
  {Celanovic}}]{chanPortableMesoscaleThermophotovoltaic2018}%
  \BibitemOpen
  \bibfield  {author} {\bibinfo {author} {\bibfnamefont {W.~R.}\ \bibnamefont
  {Chan}}, \bibinfo {author} {\bibfnamefont {V.}~\bibnamefont {Stelmakh}},
  \bibinfo {author} {\bibfnamefont {S.}~\bibnamefont {Karnani}}, \bibinfo
  {author} {\bibfnamefont {C.~M.}\ \bibnamefont {Waits}}, \bibinfo {author}
  {\bibfnamefont {M.}~\bibnamefont {Soljacic}}, \bibinfo {author}
  {\bibfnamefont {J.~D.}\ \bibnamefont {Joannopoulos}},\ and\ \bibinfo {author}
  {\bibfnamefont {I.}~\bibnamefont {Celanovic}},\ }\href
  {https://doi.org/10.1088/1742-6596/1052/1/012041} {\bibfield  {journal}
  {\bibinfo  {journal} {Journal of Physics: Conference Series}\ }\textbf
  {\bibinfo {volume} {1052}},\ \bibinfo {pages} {012041} (\bibinfo {year}
  {2018})}\BibitemShut {NoStop}%
\bibitem [{\citenamefont {Wang}\ \emph {et~al.}(2020)\citenamefont {Wang},
  \citenamefont {Liang}, \citenamefont {Fisher}, \citenamefont {Chan},\ and\
  \citenamefont {Xu}}]{wangRadioisotopeThermophotovoltaicGenerator2020}%
  \BibitemOpen
  \bibfield  {author} {\bibinfo {author} {\bibfnamefont {X.}~\bibnamefont
  {Wang}}, \bibinfo {author} {\bibfnamefont {R.}~\bibnamefont {Liang}},
  \bibinfo {author} {\bibfnamefont {P.}~\bibnamefont {Fisher}}, \bibinfo
  {author} {\bibfnamefont {W.}~\bibnamefont {Chan}},\ and\ \bibinfo {author}
  {\bibfnamefont {J.}~\bibnamefont {Xu}},\ }\href
  {https://doi.org/10.2514/1.B37623} {\bibfield  {journal} {\bibinfo  {journal}
  {Journal of Propulsion and Power}\ }\textbf {\bibinfo {volume} {36}},\
  \bibinfo {pages} {593} (\bibinfo {year} {2020})}\BibitemShut {NoStop}%
\bibitem [{\citenamefont {Rizzo}\ \emph {et~al.}(2020)\citenamefont {Rizzo},
  \citenamefont {Federici}, \citenamefont {Gatley}, \citenamefont {Gatley},
  \citenamefont {Zunino},\ and\ \citenamefont
  {Duncan}}]{rizzoComparisonTerahertzMicrowave2020a}%
  \BibitemOpen
  \bibfield  {author} {\bibinfo {author} {\bibfnamefont {L.}~\bibnamefont
  {Rizzo}}, \bibinfo {author} {\bibfnamefont {J.~F.}\ \bibnamefont {Federici}},
  \bibinfo {author} {\bibfnamefont {S.}~\bibnamefont {Gatley}}, \bibinfo
  {author} {\bibfnamefont {I.}~\bibnamefont {Gatley}}, \bibinfo {author}
  {\bibfnamefont {J.~L.}\ \bibnamefont {Zunino}},\ and\ \bibinfo {author}
  {\bibfnamefont {K.~J.}\ \bibnamefont {Duncan}},\ }\href
  {https://doi.org/10.1007/s10762-020-00719-w} {\bibfield  {journal} {\bibinfo
  {journal} {Journal of Infrared, Millimeter, and Terahertz Waves}\ }\textbf
  {\bibinfo {volume} {41}},\ \bibinfo {pages} {979} (\bibinfo {year}
  {2020})}\BibitemShut {NoStop}%
\bibitem [{\citenamefont {Yang}\ \emph {et~al.}(2022)\citenamefont {Yang},
  \citenamefont {Huang},\ and\ \citenamefont
  {Santos}}]{yangNarrowBandgapPhotovoltaic2022}%
  \BibitemOpen
  \bibfield  {author} {\bibinfo {author} {\bibfnamefont {R.~Q.}\ \bibnamefont
  {Yang}}, \bibinfo {author} {\bibfnamefont {W.}~\bibnamefont {Huang}},\ and\
  \bibinfo {author} {\bibfnamefont {M.~B.}\ \bibnamefont {Santos}},\ }\href
  {https://doi.org/10.1016/j.solmat.2022.111636} {\bibfield  {journal}
  {\bibinfo  {journal} {Solar Energy Materials and Solar Cells}\ }\textbf
  {\bibinfo {volume} {238}},\ \bibinfo {pages} {111636} (\bibinfo {year}
  {2022})}\BibitemShut {NoStop}%
\bibitem [{\citenamefont {Johnson}\ \emph {et~al.}(2008)\citenamefont
  {Johnson}, \citenamefont {Choate},\ and\ \citenamefont
  {Davidson}}]{johnsonWasteHeatRecovery2008}%
  \BibitemOpen
  \bibfield  {author} {\bibinfo {author} {\bibfnamefont {I.}~\bibnamefont
  {Johnson}}, \bibinfo {author} {\bibfnamefont {W.~T.}\ \bibnamefont
  {Choate}},\ and\ \bibinfo {author} {\bibfnamefont {A.}~\bibnamefont
  {Davidson}},\ }\href {https://doi.org/10.2172/1218716} {\emph {\bibinfo
  {title} {Waste {{Heat Recovery}}. {{Technology}} and {{Opportunities}} in
  {{U}}.{{S}}. {{Industry}}}}},\ \bibinfo {type} {Tech. Rep.}\ (\bibinfo
  {institution} {U.S. Department of Energy},\ \bibinfo {year}
  {2008})\BibitemShut {NoStop}%
\bibitem [{\citenamefont {Thekdi}\ and\ \citenamefont
  {Nimbalkar}(2015)}]{thekdiIndustrialWasteHeat2015}%
  \BibitemOpen
  \bibfield  {author} {\bibinfo {author} {\bibfnamefont {A.}~\bibnamefont
  {Thekdi}}\ and\ \bibinfo {author} {\bibfnamefont {S.~U.}\ \bibnamefont
  {Nimbalkar}},\ }\href {https://doi.org/10.2172/1185778} {\emph {\bibinfo
  {title} {Industrial {{Waste Heat Recovery}} - {{Potential Applications}},
  {{Available Technologies}} and {{Crosscutting R}}\&{{D Opportunities}}}}},\
  \bibinfo {type} {Tech. Rep.}\ (\bibinfo  {institution} {U.S. Department of
  Energy},\ \bibinfo {year} {2015})\BibitemShut {NoStop}%
\bibitem [{\citenamefont {Papapetrou}\ and\ \citenamefont
  {Kosmadakis}(2022)}]{Papapetrou2022319}%
  \BibitemOpen
  \bibfield  {author} {\bibinfo {author} {\bibfnamefont {M.}~\bibnamefont
  {Papapetrou}}\ and\ \bibinfo {author} {\bibfnamefont {G.}~\bibnamefont
  {Kosmadakis}},\ }in\ \href
  {https://doi.org/https://doi.org/10.1016/B978-0-08-102847-6.00006-1} {\emph
  {\bibinfo {booktitle} {Salinity Gradient Heat Engines}}},\ \bibinfo {series
  and number} {Woodhead Publishing Series in Energy},\ \bibinfo {editor}
  {edited by\ \bibinfo {editor} {\bibfnamefont {A.}~\bibnamefont {Tamburini}},
  \bibinfo {editor} {\bibfnamefont {A.}~\bibnamefont {Cipollina}},\ and\
  \bibinfo {editor} {\bibfnamefont {G.}~\bibnamefont {Micale}}}\ (\bibinfo
  {publisher} {Woodhead Publishing},\ \bibinfo {year} {2022})\ pp.\ \bibinfo
  {pages} {319--353}\BibitemShut {NoStop}%
\bibitem [{\citenamefont {Gamel}\ \emph
  {et~al.}(2021{\natexlab{a}})\citenamefont {Gamel}, \citenamefont {Lee},
  \citenamefont {Rashid}, \citenamefont {Ker}, \citenamefont {Yau},
  \citenamefont {Hannan},\ and\ \citenamefont
  {Jamaludin}}]{gamelReviewThermophotovoltaicCell2021}%
  \BibitemOpen
  \bibfield  {author} {\bibinfo {author} {\bibfnamefont {M.~M.~A.}\
  \bibnamefont {Gamel}}, \bibinfo {author} {\bibfnamefont {H.~J.}\ \bibnamefont
  {Lee}}, \bibinfo {author} {\bibfnamefont {W.~E. S. W.~A.}\ \bibnamefont
  {Rashid}}, \bibinfo {author} {\bibfnamefont {P.~J.}\ \bibnamefont {Ker}},
  \bibinfo {author} {\bibfnamefont {L.~K.}\ \bibnamefont {Yau}}, \bibinfo
  {author} {\bibfnamefont {M.~A.}\ \bibnamefont {Hannan}},\ and\ \bibinfo
  {author} {\bibfnamefont {M.~Z.}\ \bibnamefont {Jamaludin}},\ }\href
  {https://doi.org/10.3390/ma14174944} {\bibfield  {journal} {\bibinfo
  {journal} {Materials}\ }\textbf {\bibinfo {volume} {14}},\ \bibinfo {pages}
  {4944} (\bibinfo {year} {2021}{\natexlab{a}})}\BibitemShut {NoStop}%
\bibitem [{\citenamefont {Forcade}\ \emph {et~al.}(2022)\citenamefont
  {Forcade}, \citenamefont {Valdivia}, \citenamefont {Molesky}, \citenamefont
  {Lu}, \citenamefont {Rodriguez}, \citenamefont {Krich}, \citenamefont
  {St-Gelais},\ and\ \citenamefont {Hinzer}}]{nftpv2022}%
  \BibitemOpen
  \bibfield  {author} {\bibinfo {author} {\bibfnamefont {G.~P.}\ \bibnamefont
  {Forcade}}, \bibinfo {author} {\bibfnamefont {C.~E.}\ \bibnamefont
  {Valdivia}}, \bibinfo {author} {\bibfnamefont {S.}~\bibnamefont {Molesky}},
  \bibinfo {author} {\bibfnamefont {S.}~\bibnamefont {Lu}}, \bibinfo {author}
  {\bibfnamefont {A.~W.}\ \bibnamefont {Rodriguez}}, \bibinfo {author}
  {\bibfnamefont {J.~J.}\ \bibnamefont {Krich}}, \bibinfo {author}
  {\bibfnamefont {R.}~\bibnamefont {St-Gelais}},\ and\ \bibinfo {author}
  {\bibfnamefont {K.}~\bibnamefont {Hinzer}},\ }\href
  {https://doi.org/10.1063/5.0116806} {\bibfield  {journal} {\bibinfo
  {journal} {Applied Physics Letters}\ }\textbf {\bibinfo {volume} {121}},\
  \bibinfo {pages} {193903} (\bibinfo {year} {2022})},\ \Eprint
  {https://arxiv.org/abs/https://doi.org/10.1063/5.0116806}
  {https://doi.org/10.1063/5.0116806} \BibitemShut {NoStop}%
\bibitem [{\citenamefont {Mialhe}\ \emph {et~al.}(1900)\citenamefont {Mialhe},
  \citenamefont {Affour}, \citenamefont {El-Hajj},\ and\ \citenamefont
  {Khoury}}]{Mialhe1900}%
  \BibitemOpen
  \bibfield  {author} {\bibinfo {author} {\bibfnamefont {P.}~\bibnamefont
  {Mialhe}}, \bibinfo {author} {\bibfnamefont {B.}~\bibnamefont {Affour}},
  \bibinfo {author} {\bibfnamefont {K.}~\bibnamefont {El-Hajj}},\ and\ \bibinfo
  {author} {\bibfnamefont {A.}~\bibnamefont {Khoury}},\ }\href
  {https://doi.org/10.1155/1995/93424} {\bibfield  {journal} {\bibinfo
  {journal} {Active and Passive Electronic Components}\ }\textbf {\bibinfo
  {volume} {17}},\ \bibinfo {pages} {093424} (\bibinfo {year}
  {1900})}\BibitemShut {NoStop}%
\bibitem [{\citenamefont {Mauk}\ and\ \citenamefont
  {Andreev}(2003)}]{maukGaSbrelatedMaterialsTPV2003}%
  \BibitemOpen
  \bibfield  {author} {\bibinfo {author} {\bibfnamefont {M.~G.}\ \bibnamefont
  {Mauk}}\ and\ \bibinfo {author} {\bibfnamefont {V.~M.}\ \bibnamefont
  {Andreev}},\ }\href {https://doi.org/10.1088/0268-1242/18/5/308} {\bibfield
  {journal} {\bibinfo  {journal} {Semiconductor Science and Technology}\
  }\textbf {\bibinfo {volume} {18}},\ \bibinfo {pages} {S191} (\bibinfo {year}
  {2003})}\BibitemShut {NoStop}%
\bibitem [{\citenamefont {Wang}\ \emph {et~al.}(2003)\citenamefont {Wang},
  \citenamefont {Choi}, \citenamefont {Ransom}, \citenamefont {Charache},
  \citenamefont {Danielson},\ and\ \citenamefont
  {DePoy}}]{wangHighquantumefficiencyEVGaInAsSb1999}%
  \BibitemOpen
  \bibfield  {author} {\bibinfo {author} {\bibfnamefont {C.~A.}\ \bibnamefont
  {Wang}}, \bibinfo {author} {\bibfnamefont {H.~K.}\ \bibnamefont {Choi}},
  \bibinfo {author} {\bibfnamefont {S.~L.}\ \bibnamefont {Ransom}}, \bibinfo
  {author} {\bibfnamefont {G.~W.}\ \bibnamefont {Charache}}, \bibinfo {author}
  {\bibfnamefont {L.~R.}\ \bibnamefont {Danielson}},\ and\ \bibinfo {author}
  {\bibfnamefont {D.~M.}\ \bibnamefont {DePoy}},\ }\href
  {https://doi.org/10.1063/1.124676} {\bibfield  {journal} {\bibinfo  {journal}
  {Applied Physics Letters}\ }\textbf {\bibinfo {volume} {75}},\ \bibinfo
  {pages} {1305} (\bibinfo {year} {2003})}\BibitemShut {NoStop}%
\bibitem [{\citenamefont {Sulima}\ and\ \citenamefont
  {Bett}(2001)}]{sulimaFabricationSimulationGaSb2001a}%
  \BibitemOpen
  \bibfield  {author} {\bibinfo {author} {\bibfnamefont {O.}~\bibnamefont
  {Sulima}}\ and\ \bibinfo {author} {\bibfnamefont {A.}~\bibnamefont {Bett}},\
  }\href {https://doi.org/10.1016/S0927-0248(00)00235-X} {\bibfield  {journal}
  {\bibinfo  {journal} {Solar Energy Materials and Solar Cells}\ }\textbf
  {\bibinfo {volume} {66}},\ \bibinfo {pages} {533} (\bibinfo {year}
  {2001})}\BibitemShut {NoStop}%
\bibitem [{\citenamefont {Schlegl}(2004)}]{schleglTPVModulesBased2004}%
  \BibitemOpen
  \bibfield  {author} {\bibinfo {author} {\bibfnamefont {T.}~\bibnamefont
  {Schlegl}},\ }in\ \href {https://doi.org/10.1063/1.1841905} {\emph {\bibinfo
  {booktitle} {{{AIP Conference Proceedings}}}}},\ Vol.\ \bibinfo {volume}
  {738}\ (\bibinfo  {publisher} {{AIP}},\ \bibinfo {year} {2004})\ pp.\
  \bibinfo {pages} {285--293}\BibitemShut {NoStop}%
\bibitem [{\citenamefont {Tan}\ \emph {et~al.}(2014)\citenamefont {Tan},
  \citenamefont {Ji}, \citenamefont {Wu}, \citenamefont {Dai}, \citenamefont
  {Wang}, \citenamefont {Li}, \citenamefont {Yu}, \citenamefont {Yu},
  \citenamefont {Lu},\ and\ \citenamefont
  {Yang}}]{tanInvestigationInGaAsThermophotovoltaic2014}%
  \BibitemOpen
  \bibfield  {author} {\bibinfo {author} {\bibfnamefont {M.}~\bibnamefont
  {Tan}}, \bibinfo {author} {\bibfnamefont {L.}~\bibnamefont {Ji}}, \bibinfo
  {author} {\bibfnamefont {Y.}~\bibnamefont {Wu}}, \bibinfo {author}
  {\bibfnamefont {P.}~\bibnamefont {Dai}}, \bibinfo {author} {\bibfnamefont
  {Q.}~\bibnamefont {Wang}}, \bibinfo {author} {\bibfnamefont {K.}~\bibnamefont
  {Li}}, \bibinfo {author} {\bibfnamefont {T.}~\bibnamefont {Yu}}, \bibinfo
  {author} {\bibfnamefont {Y.}~\bibnamefont {Yu}}, \bibinfo {author}
  {\bibfnamefont {S.}~\bibnamefont {Lu}},\ and\ \bibinfo {author}
  {\bibfnamefont {H.}~\bibnamefont {Yang}},\ }\href
  {https://doi.org/10.7567/APEX.7.096601} {\bibfield  {journal} {\bibinfo
  {journal} {Applied Physics Express}\ }\textbf {\bibinfo {volume} {7}},\
  \bibinfo {pages} {096601} (\bibinfo {year} {2014})}\BibitemShut {NoStop}%
\bibitem [{\citenamefont {Gamel}\ \emph
  {et~al.}(2021{\natexlab{b}})\citenamefont {Gamel}, \citenamefont {Ker},
  \citenamefont {Rashid}, \citenamefont {Lee}, \citenamefont {Hannan},\ and\
  \citenamefont {Jamaludin}}]{gamelPerformanceGe532021}%
  \BibitemOpen
  \bibfield  {author} {\bibinfo {author} {\bibfnamefont {M.~M.~A.}\
  \bibnamefont {Gamel}}, \bibinfo {author} {\bibfnamefont {P.~J.}\ \bibnamefont
  {Ker}}, \bibinfo {author} {\bibfnamefont {W.~E. S. W.~A.}\ \bibnamefont
  {Rashid}}, \bibinfo {author} {\bibfnamefont {H.~J.}\ \bibnamefont {Lee}},
  \bibinfo {author} {\bibfnamefont {M.~A.}\ \bibnamefont {Hannan}},\ and\
  \bibinfo {author} {\bibfnamefont {M.~Z.~B.}\ \bibnamefont {Jamaludin}},\
  }\href {https://doi.org/10.1109/ACCESS.2021.3062075} {\bibfield  {journal}
  {\bibinfo  {journal} {IEEE Access}\ }\textbf {\bibinfo {volume} {9}},\
  \bibinfo {pages} {37091} (\bibinfo {year} {2021}{\natexlab{b}})}\BibitemShut
  {NoStop}%
\bibitem [{\citenamefont {Krier}\ \emph {et~al.}(2015)\citenamefont {Krier},
  \citenamefont {Yin}, \citenamefont {Marshall}, \citenamefont {Kesaria},
  \citenamefont {Krier}, \citenamefont {McDougall}, \citenamefont {Meredith},
  \citenamefont {Johnson}, \citenamefont {Inskip},\ and\ \citenamefont
  {Scholes}}]{krierLowBandgapMidinfrared2015}%
  \BibitemOpen
  \bibfield  {author} {\bibinfo {author} {\bibfnamefont {A.}~\bibnamefont
  {Krier}}, \bibinfo {author} {\bibfnamefont {M.}~\bibnamefont {Yin}}, \bibinfo
  {author} {\bibfnamefont {A.~R.~J.}\ \bibnamefont {Marshall}}, \bibinfo
  {author} {\bibfnamefont {M.}~\bibnamefont {Kesaria}}, \bibinfo {author}
  {\bibfnamefont {S.~E.}\ \bibnamefont {Krier}}, \bibinfo {author}
  {\bibfnamefont {S.}~\bibnamefont {McDougall}}, \bibinfo {author}
  {\bibfnamefont {W.}~\bibnamefont {Meredith}}, \bibinfo {author}
  {\bibfnamefont {A.~D.}\ \bibnamefont {Johnson}}, \bibinfo {author}
  {\bibfnamefont {J.}~\bibnamefont {Inskip}},\ and\ \bibinfo {author}
  {\bibfnamefont {A.}~\bibnamefont {Scholes}},\ }\href
  {https://doi.org/10.1016/j.infrared.2015.09.011} {\bibfield  {journal}
  {\bibinfo  {journal} {Infrared Physics \& Technology}\ }\textbf {\bibinfo
  {volume} {73}},\ \bibinfo {pages} {126} (\bibinfo {year} {2015})}\BibitemShut
  {NoStop}%
\bibitem [{\citenamefont {Yang}\ \emph {et~al.}(2010)\citenamefont {Yang},
  \citenamefont {Tian}, \citenamefont {Klem}, \citenamefont {Mishima},
  \citenamefont {Santos},\ and\ \citenamefont
  {Johnson}}]{yangInterbandCascadePhotovoltaic2010}%
  \BibitemOpen
  \bibfield  {author} {\bibinfo {author} {\bibfnamefont {R.~Q.}\ \bibnamefont
  {Yang}}, \bibinfo {author} {\bibfnamefont {Z.}~\bibnamefont {Tian}}, \bibinfo
  {author} {\bibfnamefont {J.~F.}\ \bibnamefont {Klem}}, \bibinfo {author}
  {\bibfnamefont {T.~D.}\ \bibnamefont {Mishima}}, \bibinfo {author}
  {\bibfnamefont {M.~B.}\ \bibnamefont {Santos}},\ and\ \bibinfo {author}
  {\bibfnamefont {M.~B.}\ \bibnamefont {Johnson}},\ }\href
  {https://doi.org/10.1063/1.3313934} {\bibfield  {journal} {\bibinfo
  {journal} {Applied Physics Letters}\ }\textbf {\bibinfo {volume} {96}},\
  \bibinfo {pages} {063504} (\bibinfo {year} {2010})}\BibitemShut {NoStop}%
\bibitem [{\citenamefont {Hinkey}\ \emph {et~al.}(2013)\citenamefont {Hinkey},
  \citenamefont {Tian}, \citenamefont {Rassel}, \citenamefont {Yang},
  \citenamefont {Klem},\ and\ \citenamefont
  {Johnson}}]{hinkeyInterbandCascadePhotovoltaic2013}%
  \BibitemOpen
  \bibfield  {author} {\bibinfo {author} {\bibfnamefont {R.~T.}\ \bibnamefont
  {Hinkey}}, \bibinfo {author} {\bibfnamefont {Z.-B.}\ \bibnamefont {Tian}},
  \bibinfo {author} {\bibfnamefont {S.~M. S.~S.}\ \bibnamefont {Rassel}},
  \bibinfo {author} {\bibfnamefont {R.~Q.}\ \bibnamefont {Yang}}, \bibinfo
  {author} {\bibfnamefont {J.~F.}\ \bibnamefont {Klem}},\ and\ \bibinfo
  {author} {\bibfnamefont {M.~B.}\ \bibnamefont {Johnson}},\ }\href
  {https://doi.org/10.1109/JPHOTOV.2013.2239360} {\bibfield  {journal}
  {\bibinfo  {journal} {IEEE Journal of Photovoltaics}\ }\textbf {\bibinfo
  {volume} {3}},\ \bibinfo {pages} {745} (\bibinfo {year} {2013})}\BibitemShut
  {NoStop}%
\bibitem [{\citenamefont {Moutanabbir}\ \emph {et~al.}(2021)\citenamefont
  {Moutanabbir}, \citenamefont {Assali}, \citenamefont {Gong}, \citenamefont
  {O'Reilly}, \citenamefont {Broderick}, \citenamefont {Marzban}, \citenamefont
  {Witzens}, \citenamefont {Du}, \citenamefont {Yu}, \citenamefont {Chelnokov},
  \citenamefont {Buca},\ and\ \citenamefont
  {Nam}}]{moutanabbirMonolithicInfraredSilicon2021}%
  \BibitemOpen
  \bibfield  {author} {\bibinfo {author} {\bibfnamefont {O.}~\bibnamefont
  {Moutanabbir}}, \bibinfo {author} {\bibfnamefont {S.}~\bibnamefont {Assali}},
  \bibinfo {author} {\bibfnamefont {X.}~\bibnamefont {Gong}}, \bibinfo {author}
  {\bibfnamefont {E.}~\bibnamefont {O'Reilly}}, \bibinfo {author}
  {\bibfnamefont {C.~A.}\ \bibnamefont {Broderick}}, \bibinfo {author}
  {\bibfnamefont {B.}~\bibnamefont {Marzban}}, \bibinfo {author} {\bibfnamefont
  {J.}~\bibnamefont {Witzens}}, \bibinfo {author} {\bibfnamefont
  {W.}~\bibnamefont {Du}}, \bibinfo {author} {\bibfnamefont {S.-Q.}\
  \bibnamefont {Yu}}, \bibinfo {author} {\bibfnamefont {A.}~\bibnamefont
  {Chelnokov}}, \bibinfo {author} {\bibfnamefont {D.}~\bibnamefont {Buca}},\
  and\ \bibinfo {author} {\bibfnamefont {D.}~\bibnamefont {Nam}},\ }\href
  {https://doi.org/10.1063/5.0043511} {\bibfield  {journal} {\bibinfo
  {journal} {Applied Physics Letters}\ }\textbf {\bibinfo {volume} {118}},\
  \bibinfo {pages} {110502} (\bibinfo {year} {2021})}\BibitemShut {NoStop}%
\bibitem [{\citenamefont {Atalla}\ \emph {et~al.}(2021)\citenamefont {Atalla},
  \citenamefont {Assali}, \citenamefont {Attiaoui}, \citenamefont
  {Lemieux-Leduc}, \citenamefont {Kumar}, \citenamefont {Abdi},\ and\
  \citenamefont {Moutanabbir}}]{atallaAllGroupIV2021b}%
  \BibitemOpen
  \bibfield  {author} {\bibinfo {author} {\bibfnamefont {M.~R.~M.}\
  \bibnamefont {Atalla}}, \bibinfo {author} {\bibfnamefont {S.}~\bibnamefont
  {Assali}}, \bibinfo {author} {\bibfnamefont {A.}~\bibnamefont {Attiaoui}},
  \bibinfo {author} {\bibfnamefont {C.}~\bibnamefont {Lemieux-Leduc}}, \bibinfo
  {author} {\bibfnamefont {A.}~\bibnamefont {Kumar}}, \bibinfo {author}
  {\bibfnamefont {S.}~\bibnamefont {Abdi}},\ and\ \bibinfo {author}
  {\bibfnamefont {O.}~\bibnamefont {Moutanabbir}},\ }\href
  {https://doi.org/10.1002/adfm.202006329} {\bibfield  {journal} {\bibinfo
  {journal} {Advanced Functional Materials}\ }\textbf {\bibinfo {volume}
  {31}},\ \bibinfo {pages} {2006329} (\bibinfo {year} {2021})}\BibitemShut
  {NoStop}%
\bibitem [{\citenamefont {Atalla}\ \emph {et~al.}(2022)\citenamefont {Atalla},
  \citenamefont {Assali}, \citenamefont {Koelling}, \citenamefont {Attiaoui},\
  and\ \citenamefont {Moutanabbir}}]{atallaHighBandwidthExtendedSWIRGeSn2022a}%
  \BibitemOpen
  \bibfield  {author} {\bibinfo {author} {\bibfnamefont {M.~R.~M.}\
  \bibnamefont {Atalla}}, \bibinfo {author} {\bibfnamefont {S.}~\bibnamefont
  {Assali}}, \bibinfo {author} {\bibfnamefont {S.}~\bibnamefont {Koelling}},
  \bibinfo {author} {\bibfnamefont {A.}~\bibnamefont {Attiaoui}},\ and\
  \bibinfo {author} {\bibfnamefont {O.}~\bibnamefont {Moutanabbir}},\ }\href
  {https://doi.org/10.1021/acsphotonics.2c00260} {\bibfield  {journal}
  {\bibinfo  {journal} {ACS Photonics}\ }\textbf {\bibinfo {volume} {9}},\
  \bibinfo {pages} {1425} (\bibinfo {year} {2022})}\BibitemShut {NoStop}%
\bibitem [{\citenamefont {Buca}\ \emph {et~al.}(2022)\citenamefont {Buca},
  \citenamefont {Bjelajac}, \citenamefont {Spirito}, \citenamefont
  {Concepción}, \citenamefont {Gromovyi}, \citenamefont {Sakat}, \citenamefont
  {Lafosse}, \citenamefont {Ferlazzo}, \citenamefont {von~den Driesch},
  \citenamefont {Ikonic}, \citenamefont {Grützmacher}, \citenamefont
  {Capellini},\ and\ \citenamefont {El~Kurdi}}]{bucaRoomTemperatureLasing2022}%
  \BibitemOpen
  \bibfield  {author} {\bibinfo {author} {\bibfnamefont {D.}~\bibnamefont
  {Buca}}, \bibinfo {author} {\bibfnamefont {A.}~\bibnamefont {Bjelajac}},
  \bibinfo {author} {\bibfnamefont {D.}~\bibnamefont {Spirito}}, \bibinfo
  {author} {\bibfnamefont {O.}~\bibnamefont {Concepción}}, \bibinfo {author}
  {\bibfnamefont {M.}~\bibnamefont {Gromovyi}}, \bibinfo {author}
  {\bibfnamefont {E.}~\bibnamefont {Sakat}}, \bibinfo {author} {\bibfnamefont
  {X.}~\bibnamefont {Lafosse}}, \bibinfo {author} {\bibfnamefont
  {L.}~\bibnamefont {Ferlazzo}}, \bibinfo {author} {\bibfnamefont
  {N.}~\bibnamefont {von~den Driesch}}, \bibinfo {author} {\bibfnamefont
  {Z.}~\bibnamefont {Ikonic}}, \bibinfo {author} {\bibfnamefont
  {D.}~\bibnamefont {Grützmacher}}, \bibinfo {author} {\bibfnamefont
  {G.}~\bibnamefont {Capellini}},\ and\ \bibinfo {author} {\bibfnamefont
  {M.}~\bibnamefont {El~Kurdi}},\ }\href
  {https://doi.org/10.1002/adom.202201024} {\bibfield  {journal} {\bibinfo
  {journal} {Advanced Optical Materials}\ }\textbf {\bibinfo {volume} {10}},\
  \bibinfo {pages} {2201024} (\bibinfo {year} {2022})}\BibitemShut {NoStop}%
\bibitem [{\citenamefont {Chang}\ \emph {et~al.}(2022)\citenamefont {Chang},
  \citenamefont {Yeh}, \citenamefont {Jheng}, \citenamefont {Hsu},
  \citenamefont {Lee}, \citenamefont {Li}, \citenamefont {Cheng},\ and\
  \citenamefont {Chang}}]{changMidinfraredResonantLight2022}%
  \BibitemOpen
  \bibfield  {author} {\bibinfo {author} {\bibfnamefont {C.-Y.}\ \bibnamefont
  {Chang}}, \bibinfo {author} {\bibfnamefont {P.-L.}\ \bibnamefont {Yeh}},
  \bibinfo {author} {\bibfnamefont {Y.-T.}\ \bibnamefont {Jheng}}, \bibinfo
  {author} {\bibfnamefont {L.-Y.}\ \bibnamefont {Hsu}}, \bibinfo {author}
  {\bibfnamefont {K.-C.}\ \bibnamefont {Lee}}, \bibinfo {author} {\bibfnamefont
  {H.}~\bibnamefont {Li}}, \bibinfo {author} {\bibfnamefont {H.~H.}\
  \bibnamefont {Cheng}},\ and\ \bibinfo {author} {\bibfnamefont {G.-E.}\
  \bibnamefont {Chang}},\ }\href {https://doi.org/10.1364/PRJ.457193}
  {\bibfield  {journal} {\bibinfo  {journal} {Photonics Research}\ }\textbf
  {\bibinfo {volume} {10}},\ \bibinfo {pages} {2278} (\bibinfo {year}
  {2022})}\BibitemShut {NoStop}%
\bibitem [{\citenamefont {Chr{\'e}tien}\ \emph {et~al.}(2019)\citenamefont
  {Chr{\'e}tien}, \citenamefont {Pauc}, \citenamefont {Armand~Pilon},
  \citenamefont {Bertrand}, \citenamefont {Thai}, \citenamefont {Casiez},
  \citenamefont {Bernier}, \citenamefont {Dansas}, \citenamefont {Gergaud},
  \citenamefont {Delamadeleine}, \citenamefont {Khazaka}, \citenamefont {Sigg},
  \citenamefont {Faist}, \citenamefont {Chelnokov}, \citenamefont {Reboud},
  \citenamefont {Hartmann},\ and\ \citenamefont
  {Calvo}}]{chretienGeSnLasersCovering2019}%
  \BibitemOpen
  \bibfield  {author} {\bibinfo {author} {\bibfnamefont {J.}~\bibnamefont
  {Chr{\'e}tien}}, \bibinfo {author} {\bibfnamefont {N.}~\bibnamefont {Pauc}},
  \bibinfo {author} {\bibfnamefont {F.}~\bibnamefont {Armand~Pilon}}, \bibinfo
  {author} {\bibfnamefont {M.}~\bibnamefont {Bertrand}}, \bibinfo {author}
  {\bibfnamefont {Q.-M.}\ \bibnamefont {Thai}}, \bibinfo {author}
  {\bibfnamefont {L.}~\bibnamefont {Casiez}}, \bibinfo {author} {\bibfnamefont
  {N.}~\bibnamefont {Bernier}}, \bibinfo {author} {\bibfnamefont
  {H.}~\bibnamefont {Dansas}}, \bibinfo {author} {\bibfnamefont
  {P.}~\bibnamefont {Gergaud}}, \bibinfo {author} {\bibfnamefont
  {E.}~\bibnamefont {Delamadeleine}}, \bibinfo {author} {\bibfnamefont
  {R.}~\bibnamefont {Khazaka}}, \bibinfo {author} {\bibfnamefont
  {H.}~\bibnamefont {Sigg}}, \bibinfo {author} {\bibfnamefont {J.}~\bibnamefont
  {Faist}}, \bibinfo {author} {\bibfnamefont {A.}~\bibnamefont {Chelnokov}},
  \bibinfo {author} {\bibfnamefont {V.}~\bibnamefont {Reboud}}, \bibinfo
  {author} {\bibfnamefont {J.-M.}\ \bibnamefont {Hartmann}},\ and\ \bibinfo
  {author} {\bibfnamefont {V.}~\bibnamefont {Calvo}},\ }\href
  {https://doi.org/10.1021/acsphotonics.9b00712} {\bibfield  {journal}
  {\bibinfo  {journal} {ACS Photonics}\ }\textbf {\bibinfo {volume} {6}},\
  \bibinfo {pages} {2462} (\bibinfo {year} {2019})}\BibitemShut {NoStop}%
\bibitem [{\citenamefont {Chrétien}\ \emph {et~al.}(2022)\citenamefont
  {Chrétien}, \citenamefont {Thai}, \citenamefont {Frauenrath}, \citenamefont
  {Casiez}, \citenamefont {Chelnokov}, \citenamefont {Reboud}, \citenamefont
  {Hartmann}, \citenamefont {El~Kurdi}, \citenamefont {Pauc},\ and\
  \citenamefont {Calvo}}]{chretienRoomTemperatureOptically2022}%
  \BibitemOpen
  \bibfield  {author} {\bibinfo {author} {\bibfnamefont {J.}~\bibnamefont
  {Chrétien}}, \bibinfo {author} {\bibfnamefont {Q.~M.}\ \bibnamefont {Thai}},
  \bibinfo {author} {\bibfnamefont {M.}~\bibnamefont {Frauenrath}}, \bibinfo
  {author} {\bibfnamefont {L.}~\bibnamefont {Casiez}}, \bibinfo {author}
  {\bibfnamefont {A.}~\bibnamefont {Chelnokov}}, \bibinfo {author}
  {\bibfnamefont {V.}~\bibnamefont {Reboud}}, \bibinfo {author} {\bibfnamefont
  {J.~M.}\ \bibnamefont {Hartmann}}, \bibinfo {author} {\bibfnamefont
  {M.}~\bibnamefont {El~Kurdi}}, \bibinfo {author} {\bibfnamefont
  {N.}~\bibnamefont {Pauc}},\ and\ \bibinfo {author} {\bibfnamefont
  {V.}~\bibnamefont {Calvo}},\ }\href {https://doi.org/10.1063/5.0074478}
  {\bibfield  {journal} {\bibinfo  {journal} {Applied Physics Letters}\
  }\textbf {\bibinfo {volume} {120}},\ \bibinfo {pages} {051107} (\bibinfo
  {year} {2022})}\BibitemShut {NoStop}%
\bibitem [{\citenamefont {Elbaz}\ \emph {et~al.}(2020)\citenamefont {Elbaz},
  \citenamefont {Buca}, \citenamefont {von~den Driesch}, \citenamefont
  {Pantzas}, \citenamefont {Patriarche}, \citenamefont {Zerounian},
  \citenamefont {Herth}, \citenamefont {Checoury}, \citenamefont {Sauvage},
  \citenamefont {Sagnes}, \citenamefont {Foti}, \citenamefont {Ossikovski},
  \citenamefont {Hartmann}, \citenamefont {Boeuf}, \citenamefont {Ikonic},
  \citenamefont {Boucaud}, \citenamefont {Grützmacher},\ and\ \citenamefont
  {El~Kurdi}}]{elbazUltralowthresholdContinuouswavePulsed2020b}%
  \BibitemOpen
  \bibfield  {author} {\bibinfo {author} {\bibfnamefont {A.}~\bibnamefont
  {Elbaz}}, \bibinfo {author} {\bibfnamefont {D.}~\bibnamefont {Buca}},
  \bibinfo {author} {\bibfnamefont {N.}~\bibnamefont {von~den Driesch}},
  \bibinfo {author} {\bibfnamefont {K.}~\bibnamefont {Pantzas}}, \bibinfo
  {author} {\bibfnamefont {G.}~\bibnamefont {Patriarche}}, \bibinfo {author}
  {\bibfnamefont {N.}~\bibnamefont {Zerounian}}, \bibinfo {author}
  {\bibfnamefont {E.}~\bibnamefont {Herth}}, \bibinfo {author} {\bibfnamefont
  {X.}~\bibnamefont {Checoury}}, \bibinfo {author} {\bibfnamefont
  {S.}~\bibnamefont {Sauvage}}, \bibinfo {author} {\bibfnamefont
  {I.}~\bibnamefont {Sagnes}}, \bibinfo {author} {\bibfnamefont
  {A.}~\bibnamefont {Foti}}, \bibinfo {author} {\bibfnamefont {R.}~\bibnamefont
  {Ossikovski}}, \bibinfo {author} {\bibfnamefont {J.-M.}\ \bibnamefont
  {Hartmann}}, \bibinfo {author} {\bibfnamefont {F.}~\bibnamefont {Boeuf}},
  \bibinfo {author} {\bibfnamefont {Z.}~\bibnamefont {Ikonic}}, \bibinfo
  {author} {\bibfnamefont {P.}~\bibnamefont {Boucaud}}, \bibinfo {author}
  {\bibfnamefont {D.}~\bibnamefont {Grützmacher}},\ and\ \bibinfo {author}
  {\bibfnamefont {M.}~\bibnamefont {El~Kurdi}},\ }\href
  {https://doi.org/10.1038/s41566-020-0601-5} {\bibfield  {journal} {\bibinfo
  {journal} {Nature Photonics}\ }\textbf {\bibinfo {volume} {14}},\ \bibinfo
  {pages} {375} (\bibinfo {year} {2020})}\BibitemShut {NoStop}%
\bibitem [{\citenamefont {Joo}\ \emph {et~al.}(2021)\citenamefont {Joo},
  \citenamefont {Kim}, \citenamefont {Burt}, \citenamefont {Jung},
  \citenamefont {Zhang}, \citenamefont {Chen}, \citenamefont {Parluhutan},
  \citenamefont {Kang}, \citenamefont {Lee}, \citenamefont {Assali},
  \citenamefont {Ikonic}, \citenamefont {Moutanabbir}, \citenamefont {Cho},
  \citenamefont {Tan},\ and\ \citenamefont {Nam}}]{joo1DPhotonicCrystal2021}%
  \BibitemOpen
  \bibfield  {author} {\bibinfo {author} {\bibfnamefont {H.-J.}\ \bibnamefont
  {Joo}}, \bibinfo {author} {\bibfnamefont {Y.}~\bibnamefont {Kim}}, \bibinfo
  {author} {\bibfnamefont {D.}~\bibnamefont {Burt}}, \bibinfo {author}
  {\bibfnamefont {Y.}~\bibnamefont {Jung}}, \bibinfo {author} {\bibfnamefont
  {L.}~\bibnamefont {Zhang}}, \bibinfo {author} {\bibfnamefont
  {M.}~\bibnamefont {Chen}}, \bibinfo {author} {\bibfnamefont {S.~J.}\
  \bibnamefont {Parluhutan}}, \bibinfo {author} {\bibfnamefont {D.-H.}\
  \bibnamefont {Kang}}, \bibinfo {author} {\bibfnamefont {C.}~\bibnamefont
  {Lee}}, \bibinfo {author} {\bibfnamefont {S.}~\bibnamefont {Assali}},
  \bibinfo {author} {\bibfnamefont {Z.}~\bibnamefont {Ikonic}}, \bibinfo
  {author} {\bibfnamefont {O.}~\bibnamefont {Moutanabbir}}, \bibinfo {author}
  {\bibfnamefont {Y.-H.}\ \bibnamefont {Cho}}, \bibinfo {author} {\bibfnamefont
  {C.~S.}\ \bibnamefont {Tan}},\ and\ \bibinfo {author} {\bibfnamefont
  {D.}~\bibnamefont {Nam}},\ }\href {https://doi.org/10.1063/5.0066935}
  {\bibfield  {journal} {\bibinfo  {journal} {Applied Physics Letters}\
  }\textbf {\bibinfo {volume} {119}},\ \bibinfo {pages} {201101} (\bibinfo
  {year} {2021})}\BibitemShut {NoStop}%
\bibitem [{\citenamefont {Jung}\ \emph {et~al.}(2022)\citenamefont {Jung},
  \citenamefont {Burt}, \citenamefont {Zhang}, \citenamefont {Kim},
  \citenamefont {Joo}, \citenamefont {Chen}, \citenamefont {Assali},
  \citenamefont {Moutanabbir}, \citenamefont {Tan},\ and\ \citenamefont
  {Nam}}]{jungOpticallyPumpedLowthreshold2022}%
  \BibitemOpen
  \bibfield  {author} {\bibinfo {author} {\bibfnamefont {Y.}~\bibnamefont
  {Jung}}, \bibinfo {author} {\bibfnamefont {D.}~\bibnamefont {Burt}}, \bibinfo
  {author} {\bibfnamefont {L.}~\bibnamefont {Zhang}}, \bibinfo {author}
  {\bibfnamefont {Y.}~\bibnamefont {Kim}}, \bibinfo {author} {\bibfnamefont
  {H.-J.}\ \bibnamefont {Joo}}, \bibinfo {author} {\bibfnamefont
  {M.}~\bibnamefont {Chen}}, \bibinfo {author} {\bibfnamefont {S.}~\bibnamefont
  {Assali}}, \bibinfo {author} {\bibfnamefont {O.}~\bibnamefont {Moutanabbir}},
  \bibinfo {author} {\bibfnamefont {C.~S.}\ \bibnamefont {Tan}},\ and\ \bibinfo
  {author} {\bibfnamefont {D.}~\bibnamefont {Nam}},\ }\href
  {https://doi.org/10.1364/PRJ.455443} {\bibfield  {journal} {\bibinfo
  {journal} {Photonics Research}\ }\textbf {\bibinfo {volume} {10}},\ \bibinfo
  {pages} {1332} (\bibinfo {year} {2022})}\BibitemShut {NoStop}%
\bibitem [{\citenamefont {Li}\ \emph {et~al.}(2021)\citenamefont {Li},
  \citenamefont {Peng}, \citenamefont {Liu}, \citenamefont {Zhou},
  \citenamefont {Zheng}, \citenamefont {Xue}, \citenamefont {Zuo},
  \citenamefont {Chen},\ and\ \citenamefont {Cheng}}]{li30GHzGeSn2021}%
  \BibitemOpen
  \bibfield  {author} {\bibinfo {author} {\bibfnamefont {X.}~\bibnamefont
  {Li}}, \bibinfo {author} {\bibfnamefont {L.}~\bibnamefont {Peng}}, \bibinfo
  {author} {\bibfnamefont {Z.}~\bibnamefont {Liu}}, \bibinfo {author}
  {\bibfnamefont {Z.}~\bibnamefont {Zhou}}, \bibinfo {author} {\bibfnamefont
  {J.}~\bibnamefont {Zheng}}, \bibinfo {author} {\bibfnamefont
  {C.}~\bibnamefont {Xue}}, \bibinfo {author} {\bibfnamefont {Y.}~\bibnamefont
  {Zuo}}, \bibinfo {author} {\bibfnamefont {B.}~\bibnamefont {Chen}},\ and\
  \bibinfo {author} {\bibfnamefont {B.}~\bibnamefont {Cheng}},\ }\href
  {https://doi.org/10.1364/PRJ.413453} {\bibfield  {journal} {\bibinfo
  {journal} {Photonics Research}\ }\textbf {\bibinfo {volume} {9}},\ \bibinfo
  {pages} {494} (\bibinfo {year} {2021})}\BibitemShut {NoStop}%
\bibitem [{\citenamefont {Liu}\ \emph {et~al.}(2022)\citenamefont {Liu},
  \citenamefont {Zheng}, \citenamefont {Niu}, \citenamefont {Liu},
  \citenamefont {Huang}, \citenamefont {Li}, \citenamefont {Zhang},
  \citenamefont {Pang}, \citenamefont {Liu}, \citenamefont {Zuo},\ and\
  \citenamefont {Cheng}}]{liuSnContentGradient2022}%
  \BibitemOpen
  \bibfield  {author} {\bibinfo {author} {\bibfnamefont {X.}~\bibnamefont
  {Liu}}, \bibinfo {author} {\bibfnamefont {J.}~\bibnamefont {Zheng}}, \bibinfo
  {author} {\bibfnamefont {C.}~\bibnamefont {Niu}}, \bibinfo {author}
  {\bibfnamefont {T.}~\bibnamefont {Liu}}, \bibinfo {author} {\bibfnamefont
  {Q.}~\bibnamefont {Huang}}, \bibinfo {author} {\bibfnamefont
  {M.}~\bibnamefont {Li}}, \bibinfo {author} {\bibfnamefont {D.}~\bibnamefont
  {Zhang}}, \bibinfo {author} {\bibfnamefont {Y.}~\bibnamefont {Pang}},
  \bibinfo {author} {\bibfnamefont {Z.}~\bibnamefont {Liu}}, \bibinfo {author}
  {\bibfnamefont {Y.}~\bibnamefont {Zuo}},\ and\ \bibinfo {author}
  {\bibfnamefont {B.}~\bibnamefont {Cheng}},\ }\href
  {https://doi.org/10.1364/PRJ.456000} {\bibfield  {journal} {\bibinfo
  {journal} {Photonics Research}\ }\textbf {\bibinfo {volume} {10}},\ \bibinfo
  {pages} {1567} (\bibinfo {year} {2022})}\BibitemShut {NoStop}%
\bibitem [{\citenamefont {Luo}\ \emph {et~al.}(2022)\citenamefont {Luo},
  \citenamefont {Assali}, \citenamefont {Atalla}, \citenamefont {Koelling},
  \citenamefont {Attiaoui}, \citenamefont {Daligou}, \citenamefont {Martí},
  \citenamefont {Arbiol},\ and\ \citenamefont
  {Moutanabbir}}]{luoExtendedSWIRPhotodetectionAllGroup2022b}%
  \BibitemOpen
  \bibfield  {author} {\bibinfo {author} {\bibfnamefont {L.}~\bibnamefont
  {Luo}}, \bibinfo {author} {\bibfnamefont {S.}~\bibnamefont {Assali}},
  \bibinfo {author} {\bibfnamefont {M.~R.~M.}\ \bibnamefont {Atalla}}, \bibinfo
  {author} {\bibfnamefont {S.}~\bibnamefont {Koelling}}, \bibinfo {author}
  {\bibfnamefont {A.}~\bibnamefont {Attiaoui}}, \bibinfo {author}
  {\bibfnamefont {G.}~\bibnamefont {Daligou}}, \bibinfo {author} {\bibfnamefont
  {S.}~\bibnamefont {Martí}}, \bibinfo {author} {\bibfnamefont
  {J.}~\bibnamefont {Arbiol}},\ and\ \bibinfo {author} {\bibfnamefont
  {O.}~\bibnamefont {Moutanabbir}},\ }\href
  {https://doi.org/10.1021/acsphotonics.1c01728} {\bibfield  {journal}
  {\bibinfo  {journal} {ACS Photonics}\ }\textbf {\bibinfo {volume} {9}},\
  \bibinfo {pages} {914} (\bibinfo {year} {2022})}\BibitemShut {NoStop}%
\bibitem [{\citenamefont {Marzban}\ \emph {et~al.}(2022)\citenamefont
  {Marzban}, \citenamefont {Seidel}, \citenamefont {Liu}, \citenamefont {Wu},
  \citenamefont {Kiyek}, \citenamefont {Zoellner}, \citenamefont {Ikonic},
  \citenamefont {Schulze}, \citenamefont {Grützmacher}, \citenamefont
  {Capellini}, \citenamefont {Oehme}, \citenamefont {Witzens},\ and\
  \citenamefont {Buca}}]{marzbanStrainEngineeredElectrically2022}%
  \BibitemOpen
  \bibfield  {author} {\bibinfo {author} {\bibfnamefont {B.}~\bibnamefont
  {Marzban}}, \bibinfo {author} {\bibfnamefont {L.}~\bibnamefont {Seidel}},
  \bibinfo {author} {\bibfnamefont {T.}~\bibnamefont {Liu}}, \bibinfo {author}
  {\bibfnamefont {K.}~\bibnamefont {Wu}}, \bibinfo {author} {\bibfnamefont
  {V.}~\bibnamefont {Kiyek}}, \bibinfo {author} {\bibfnamefont {M.~H.}\
  \bibnamefont {Zoellner}}, \bibinfo {author} {\bibfnamefont {Z.}~\bibnamefont
  {Ikonic}}, \bibinfo {author} {\bibfnamefont {J.}~\bibnamefont {Schulze}},
  \bibinfo {author} {\bibfnamefont {D.}~\bibnamefont {Grützmacher}}, \bibinfo
  {author} {\bibfnamefont {G.}~\bibnamefont {Capellini}}, \bibinfo {author}
  {\bibfnamefont {M.}~\bibnamefont {Oehme}}, \bibinfo {author} {\bibfnamefont
  {J.}~\bibnamefont {Witzens}},\ and\ \bibinfo {author} {\bibfnamefont
  {D.}~\bibnamefont {Buca}},\ }\href
  {https://doi.org/10.1021/acsphotonics.2c01508} {\bibfield  {journal}
  {\bibinfo  {journal} {ACS Photonics}\ ,\ \bibinfo {pages}
  {acsphotonics.2c01508}} (\bibinfo {year} {2022})}\BibitemShut {NoStop}%
\bibitem [{\citenamefont {Talamas~Simola}\ \emph {et~al.}(2021)\citenamefont
  {Talamas~Simola}, \citenamefont {Kiyek}, \citenamefont {Ballabio},
  \citenamefont {Schlykow}, \citenamefont {Frigerio}, \citenamefont
  {Zucchetti}, \citenamefont {De~Iacovo}, \citenamefont {Colace}, \citenamefont
  {Yamamoto}, \citenamefont {Capellini}, \citenamefont {Gr\"{u}tzmacher},
  \citenamefont {Buca},\ and\ \citenamefont
  {Isella}}]{talamassimolaCMOSCompatibleBiasTunableDualBand2021}%
  \BibitemOpen
  \bibfield  {author} {\bibinfo {author} {\bibfnamefont {E.}~\bibnamefont
  {Talamas~Simola}}, \bibinfo {author} {\bibfnamefont {V.}~\bibnamefont
  {Kiyek}}, \bibinfo {author} {\bibfnamefont {A.}~\bibnamefont {Ballabio}},
  \bibinfo {author} {\bibfnamefont {V.}~\bibnamefont {Schlykow}}, \bibinfo
  {author} {\bibfnamefont {J.}~\bibnamefont {Frigerio}}, \bibinfo {author}
  {\bibfnamefont {C.}~\bibnamefont {Zucchetti}}, \bibinfo {author}
  {\bibfnamefont {A.}~\bibnamefont {De~Iacovo}}, \bibinfo {author}
  {\bibfnamefont {L.}~\bibnamefont {Colace}}, \bibinfo {author} {\bibfnamefont
  {Y.}~\bibnamefont {Yamamoto}}, \bibinfo {author} {\bibfnamefont
  {G.}~\bibnamefont {Capellini}}, \bibinfo {author} {\bibfnamefont
  {D.}~\bibnamefont {Gr\"{u}tzmacher}}, \bibinfo {author} {\bibfnamefont
  {D.}~\bibnamefont {Buca}},\ and\ \bibinfo {author} {\bibfnamefont
  {G.}~\bibnamefont {Isella}},\ }\href
  {https://doi.org/10.1021/acsphotonics.1c00617} {\bibfield  {journal}
  {\bibinfo  {journal} {ACS Photonics}\ }\textbf {\bibinfo {volume} {8}},\
  \bibinfo {pages} {2166} (\bibinfo {year} {2021})}\BibitemShut {NoStop}%
\bibitem [{\citenamefont {Tran}\ \emph {et~al.}(2019)\citenamefont {Tran},
  \citenamefont {Pham}, \citenamefont {Margetis}, \citenamefont {Zhou},
  \citenamefont {Dou}, \citenamefont {Grant}, \citenamefont {Grant},
  \citenamefont {Al-Kabi}, \citenamefont {Sun}, \citenamefont {Soref},
  \citenamefont {Tolle}, \citenamefont {Zhang}, \citenamefont {Du},
  \citenamefont {Li}, \citenamefont {Mortazavi},\ and\ \citenamefont
  {Yu}}]{tranSiBasedGeSnPhotodetectors2019}%
  \BibitemOpen
  \bibfield  {author} {\bibinfo {author} {\bibfnamefont {H.}~\bibnamefont
  {Tran}}, \bibinfo {author} {\bibfnamefont {T.}~\bibnamefont {Pham}}, \bibinfo
  {author} {\bibfnamefont {J.}~\bibnamefont {Margetis}}, \bibinfo {author}
  {\bibfnamefont {Y.}~\bibnamefont {Zhou}}, \bibinfo {author} {\bibfnamefont
  {W.}~\bibnamefont {Dou}}, \bibinfo {author} {\bibfnamefont {P.~C.}\
  \bibnamefont {Grant}}, \bibinfo {author} {\bibfnamefont {J.~M.}\ \bibnamefont
  {Grant}}, \bibinfo {author} {\bibfnamefont {S.}~\bibnamefont {Al-Kabi}},
  \bibinfo {author} {\bibfnamefont {G.}~\bibnamefont {Sun}}, \bibinfo {author}
  {\bibfnamefont {R.~A.}\ \bibnamefont {Soref}}, \bibinfo {author}
  {\bibfnamefont {J.}~\bibnamefont {Tolle}}, \bibinfo {author} {\bibfnamefont
  {Y.-H.}\ \bibnamefont {Zhang}}, \bibinfo {author} {\bibfnamefont
  {W.}~\bibnamefont {Du}}, \bibinfo {author} {\bibfnamefont {B.}~\bibnamefont
  {Li}}, \bibinfo {author} {\bibfnamefont {M.}~\bibnamefont {Mortazavi}},\ and\
  \bibinfo {author} {\bibfnamefont {S.-Q.}\ \bibnamefont {Yu}},\ }\href
  {https://doi.org/10.1021/acsphotonics.9b00845} {\bibfield  {journal}
  {\bibinfo  {journal} {ACS Photonics}\ }\textbf {\bibinfo {volume} {6}},\
  \bibinfo {pages} {2807} (\bibinfo {year} {2019})}\BibitemShut {NoStop}%
\bibitem [{\citenamefont {Xu}\ \emph {et~al.}(2019)\citenamefont {Xu},
  \citenamefont {Wang}, \citenamefont {Huang}, \citenamefont {Dong},
  \citenamefont {{Masudy-Panah}}, \citenamefont {Wang}, \citenamefont {Gong},\
  and\ \citenamefont {Yeo}}]{xuHighspeedPhotoDetection2019}%
  \BibitemOpen
  \bibfield  {author} {\bibinfo {author} {\bibfnamefont {S.}~\bibnamefont
  {Xu}}, \bibinfo {author} {\bibfnamefont {W.}~\bibnamefont {Wang}}, \bibinfo
  {author} {\bibfnamefont {Y.-C.}\ \bibnamefont {Huang}}, \bibinfo {author}
  {\bibfnamefont {Y.}~\bibnamefont {Dong}}, \bibinfo {author} {\bibfnamefont
  {S.}~\bibnamefont {{Masudy-Panah}}}, \bibinfo {author} {\bibfnamefont
  {H.}~\bibnamefont {Wang}}, \bibinfo {author} {\bibfnamefont {X.}~\bibnamefont
  {Gong}},\ and\ \bibinfo {author} {\bibfnamefont {Y.-C.}\ \bibnamefont
  {Yeo}},\ }\href {https://doi.org/10.1364/OE.27.005798} {\bibfield  {journal}
  {\bibinfo  {journal} {Optics Express}\ }\textbf {\bibinfo {volume} {27}},\
  \bibinfo {pages} {5798} (\bibinfo {year} {2019})}\BibitemShut {NoStop}%
\bibitem [{\citenamefont {Zhou}\ \emph {et~al.}(2020)\citenamefont {Zhou},
  \citenamefont {Miao}, \citenamefont {Ojo}, \citenamefont {Tran},
  \citenamefont {Abernathy}, \citenamefont {Grant}, \citenamefont {Amoah},
  \citenamefont {Salamo}, \citenamefont {Du}, \citenamefont {Liu},
  \citenamefont {Margetis}, \citenamefont {Tolle}, \citenamefont {Zhang},
  \citenamefont {Sun}, \citenamefont {Soref}, \citenamefont {Li},\ and\
  \citenamefont {Yu}}]{zhouElectricallyInjectedGeSn2020a}%
  \BibitemOpen
  \bibfield  {author} {\bibinfo {author} {\bibfnamefont {Y.}~\bibnamefont
  {Zhou}}, \bibinfo {author} {\bibfnamefont {Y.}~\bibnamefont {Miao}}, \bibinfo
  {author} {\bibfnamefont {S.}~\bibnamefont {Ojo}}, \bibinfo {author}
  {\bibfnamefont {H.}~\bibnamefont {Tran}}, \bibinfo {author} {\bibfnamefont
  {G.}~\bibnamefont {Abernathy}}, \bibinfo {author} {\bibfnamefont {J.~M.}\
  \bibnamefont {Grant}}, \bibinfo {author} {\bibfnamefont {S.}~\bibnamefont
  {Amoah}}, \bibinfo {author} {\bibfnamefont {G.}~\bibnamefont {Salamo}},
  \bibinfo {author} {\bibfnamefont {W.}~\bibnamefont {Du}}, \bibinfo {author}
  {\bibfnamefont {J.}~\bibnamefont {Liu}}, \bibinfo {author} {\bibfnamefont
  {J.}~\bibnamefont {Margetis}}, \bibinfo {author} {\bibfnamefont
  {J.}~\bibnamefont {Tolle}}, \bibinfo {author} {\bibfnamefont {Y.-h.}\
  \bibnamefont {Zhang}}, \bibinfo {author} {\bibfnamefont {G.}~\bibnamefont
  {Sun}}, \bibinfo {author} {\bibfnamefont {R.~A.}\ \bibnamefont {Soref}},
  \bibinfo {author} {\bibfnamefont {B.}~\bibnamefont {Li}},\ and\ \bibinfo
  {author} {\bibfnamefont {S.-Q.}\ \bibnamefont {Yu}},\ }\href
  {https://doi.org/10.1364/OPTICA.395687} {\bibfield  {journal} {\bibinfo
  {journal} {Optica}\ }\textbf {\bibinfo {volume} {7}},\ \bibinfo {pages} {924}
  (\bibinfo {year} {2020})}\BibitemShut {NoStop}%
\bibitem [{\citenamefont {Nelson}(2003)}]{Nelson2003}%
  \BibitemOpen
  \bibfield  {author} {\bibinfo {author} {\bibfnamefont {J.}~\bibnamefont
  {Nelson}},\ }\href {https://doi.org/10.1142/p276} {\emph {\bibinfo {title}
  {The Physics of Solar Cells}}}\ (\bibinfo  {publisher} {Imperial College
  Press},\ \bibinfo {year} {2003})\BibitemShut {NoStop}%
\bibitem [{\citenamefont {Sze}\ and\ \citenamefont
  {Ng}(2007)}]{szePhysicsSemiconductorDevices2007}%
  \BibitemOpen
  \bibfield  {author} {\bibinfo {author} {\bibfnamefont {S.~M.}\ \bibnamefont
  {Sze}}\ and\ \bibinfo {author} {\bibfnamefont {K.~K.}\ \bibnamefont {Ng}},\
  }\href@noop {} {\emph {\bibinfo {title} {Physics of Semiconductor
  Devices}}},\ \bibinfo {edition} {3rd}\ ed.\ (\bibinfo  {publisher}
  {{Wiley-Interscience}},\ \bibinfo {year} {2007})\BibitemShut {NoStop}%
\bibitem [{\citenamefont {Pettersson}\ \emph {et~al.}(1999)\citenamefont
  {Pettersson}, \citenamefont {Roman},\ and\ \citenamefont
  {Ingan\"{a}s}}]{petterssonModelingPhotocurrentAction1999}%
  \BibitemOpen
  \bibfield  {author} {\bibinfo {author} {\bibfnamefont {L.~A.~A.}\
  \bibnamefont {Pettersson}}, \bibinfo {author} {\bibfnamefont {L.~S.}\
  \bibnamefont {Roman}},\ and\ \bibinfo {author} {\bibfnamefont
  {O.}~\bibnamefont {Ingan\"{a}s}},\ }\href {https://doi.org/10.1063/1.370757}
  {\bibfield  {journal} {\bibinfo  {journal} {Journal of Applied Physics}\
  }\textbf {\bibinfo {volume} {86}},\ \bibinfo {pages} {487} (\bibinfo {year}
  {1999})}\BibitemShut {NoStop}%
\bibitem [{\citenamefont {Donges}(1998)}]{AxelDonges1998}%
  \BibitemOpen
  \bibfield  {author} {\bibinfo {author} {\bibfnamefont {A.}~\bibnamefont
  {Donges}},\ }\href {https://doi.org/10.1088/0143-0807/19/3/006} {\bibfield
  {journal} {\bibinfo  {journal} {European Journal of Physics}\ }\textbf
  {\bibinfo {volume} {19}},\ \bibinfo {pages} {245} (\bibinfo {year}
  {1998})}\BibitemShut {NoStop}%
\bibitem [{\citenamefont {Katsidis}\ and\ \citenamefont
  {Siapkas}(2002)}]{Katsidis02}%
  \BibitemOpen
  \bibfield  {author} {\bibinfo {author} {\bibfnamefont {C.~C.}\ \bibnamefont
  {Katsidis}}\ and\ \bibinfo {author} {\bibfnamefont {D.~I.}\ \bibnamefont
  {Siapkas}},\ }\href {https://doi.org/10.1364/AO.41.003978} {\bibfield
  {journal} {\bibinfo  {journal} {Appl. Opt.}\ }\textbf {\bibinfo {volume}
  {41}},\ \bibinfo {pages} {3978} (\bibinfo {year} {2002})}\BibitemShut
  {NoStop}%
\bibitem [{\citenamefont {Assali}\ \emph {et~al.}(2021)\citenamefont {Assali},
  \citenamefont {Dijkstra}, \citenamefont {Attiaoui}, \citenamefont
  {Bouthillier}, \citenamefont {Haverkort},\ and\ \citenamefont
  {Moutanabbir}}]{Assali2021}%
  \BibitemOpen
  \bibfield  {author} {\bibinfo {author} {\bibfnamefont {S.}~\bibnamefont
  {Assali}}, \bibinfo {author} {\bibfnamefont {A.}~\bibnamefont {Dijkstra}},
  \bibinfo {author} {\bibfnamefont {A.}~\bibnamefont {Attiaoui}}, \bibinfo
  {author} {\bibfnamefont {E.}~\bibnamefont {Bouthillier}}, \bibinfo {author}
  {\bibfnamefont {J.}~\bibnamefont {Haverkort}},\ and\ \bibinfo {author}
  {\bibfnamefont {O.}~\bibnamefont {Moutanabbir}},\ }\href
  {https://doi.org/10.1103/PhysRevApplied.15.024031} {\bibfield  {journal}
  {\bibinfo  {journal} {Physical Review Applied}\ }\textbf {\bibinfo {volume}
  {15}},\ \bibinfo {pages} {024031} (\bibinfo {year} {2021})}\BibitemShut
  {NoStop}%
\bibitem [{\citenamefont {Chang}\ \emph {et~al.}(2010)\citenamefont {Chang},
  \citenamefont {Chang},\ and\ \citenamefont {Chuang}}]{Chang2010}%
  \BibitemOpen
  \bibfield  {author} {\bibinfo {author} {\bibfnamefont {G.-E.}\ \bibnamefont
  {Chang}}, \bibinfo {author} {\bibfnamefont {S.-w.}\ \bibnamefont {Chang}},\
  and\ \bibinfo {author} {\bibfnamefont {S.~L.}\ \bibnamefont {Chuang}},\
  }\href {https://doi.org/10.1109/JQE.2010.2059000} {\bibfield  {journal}
  {\bibinfo  {journal} {IEEE Journal of Quantum Electronics}\ }\textbf
  {\bibinfo {volume} {46}},\ \bibinfo {pages} {1813} (\bibinfo {year}
  {2010})}\BibitemShut {NoStop}%
\bibitem [{\citenamefont {Bosi}\ and\ \citenamefont
  {Attolini}(2010)}]{bosi2010}%
  \BibitemOpen
  \bibfield  {author} {\bibinfo {author} {\bibfnamefont {M.}~\bibnamefont
  {Bosi}}\ and\ \bibinfo {author} {\bibfnamefont {G.}~\bibnamefont
  {Attolini}},\ }\href {https://doi.org/10.1016/j.pcrysgrow.2010.09.002}
  {\bibfield  {journal} {\bibinfo  {journal} {Progress in Crystal Growth and
  Characterization of Materials}\ }\textbf {\bibinfo {volume} {56}},\ \bibinfo
  {pages} {146} (\bibinfo {year} {2010})}\BibitemShut {NoStop}%
\bibitem [{\citenamefont {Caughey}\ and\ \citenamefont
  {Thomas}(1967)}]{caughey1967}%
  \BibitemOpen
  \bibfield  {author} {\bibinfo {author} {\bibfnamefont {D.}~\bibnamefont
  {Caughey}}\ and\ \bibinfo {author} {\bibfnamefont {R.}~\bibnamefont
  {Thomas}},\ }\href {https://doi.org/10.1109/PROC.1967.6123} {\bibfield
  {journal} {\bibinfo  {journal} {Proceedings of the IEEE}\ }\textbf {\bibinfo
  {volume} {55}},\ \bibinfo {pages} {2192} (\bibinfo {year}
  {1967})}\BibitemShut {NoStop}%
\bibitem [{\citenamefont {Virgilio}\ \emph {et~al.}(2013)\citenamefont
  {Virgilio}, \citenamefont {Manganelli}, \citenamefont {Grosso}, \citenamefont
  {Schroeder},\ and\ \citenamefont {Capellini}}]{virgilio2013}%
  \BibitemOpen
  \bibfield  {author} {\bibinfo {author} {\bibfnamefont {M.}~\bibnamefont
  {Virgilio}}, \bibinfo {author} {\bibfnamefont {C.~L.}\ \bibnamefont
  {Manganelli}}, \bibinfo {author} {\bibfnamefont {G.}~\bibnamefont {Grosso}},
  \bibinfo {author} {\bibfnamefont {T.}~\bibnamefont {Schroeder}},\ and\
  \bibinfo {author} {\bibfnamefont {G.}~\bibnamefont {Capellini}},\ }\href
  {https://doi.org/10.1063/1.4849855} {\bibfield  {journal} {\bibinfo
  {journal} {Journal of Applied Physics}\ }\textbf {\bibinfo {volume} {114}},\
  \bibinfo {pages} {243102} (\bibinfo {year} {2013})}\BibitemShut {NoStop}%
\bibitem [{\citenamefont {Ioffe}()}]{GeAffinityIoffe}%
  \BibitemOpen
  \bibfield  {author} {\bibinfo {author} {\bibnamefont {Ioffe}},\ }\href@noop
  {} {\bibinfo {title} {Basic parameters at 300 k}},\ \bibinfo {howpublished}
  {\url{http://www.ioffe.ru/SVA/NSM/Semicond/Ge/basic.html}}\BibitemShut
  {NoStop}%
\bibitem [{\citenamefont {Conradt}\ and\ \citenamefont
  {Aengenheister}(1972)}]{conradt1972}%
  \BibitemOpen
  \bibfield  {author} {\bibinfo {author} {\bibfnamefont {R.}~\bibnamefont
  {Conradt}}\ and\ \bibinfo {author} {\bibfnamefont {J.}~\bibnamefont
  {Aengenheister}},\ }\href {https://doi.org/10.1016/0038-1098(72)90016-6}
  {\bibfield  {journal} {\bibinfo  {journal} {Solid State Communications}\
  }\textbf {\bibinfo {volume} {10}},\ \bibinfo {pages} {321} (\bibinfo {year}
  {1972})}\BibitemShut {NoStop}%
\bibitem [{\citenamefont {Marchetti}\ \emph {et~al.}(2001)\citenamefont
  {Marchetti}, \citenamefont {Martinelli}, \citenamefont {Simili},
  \citenamefont {Giorgi},\ and\ \citenamefont {Fantoni}}]{marchetti2001}%
  \BibitemOpen
  \bibfield  {author} {\bibinfo {author} {\bibfnamefont {S.}~\bibnamefont
  {Marchetti}}, \bibinfo {author} {\bibfnamefont {M.}~\bibnamefont
  {Martinelli}}, \bibinfo {author} {\bibfnamefont {R.}~\bibnamefont {Simili}},
  \bibinfo {author} {\bibfnamefont {M.}~\bibnamefont {Giorgi}},\ and\ \bibinfo
  {author} {\bibfnamefont {R.}~\bibnamefont {Fantoni}},\ }\href
  {https://doi.org/10.1238/Physica.Regular.064a00509} {\bibfield  {journal}
  {\bibinfo  {journal} {Physica Scripta}\ }\textbf {\bibinfo {volume} {64}},\
  \bibinfo {pages} {509} (\bibinfo {year} {2001})}\BibitemShut {NoStop}%
\bibitem [{\citenamefont {Chuang}(2009)}]{chuang2012physics}%
  \BibitemOpen
  \bibfield  {author} {\bibinfo {author} {\bibfnamefont {S.~L.}\ \bibnamefont
  {Chuang}},\ }\href@noop {} {\emph {\bibinfo {title} {Physics of Photonic
  Devices}}},\ \bibinfo {edition} {2nd}\ ed.,\ Wiley Series in Pure and Applied
  Optics\ (\bibinfo  {publisher} {{Wiley}},\ \bibinfo {year}
  {2009})\BibitemShut {NoStop}%
\bibitem [{\citenamefont {Aymerich-Humet}\ \emph {et~al.}(1983)\citenamefont
  {Aymerich-Humet}, \citenamefont {Serra-Mestres},\ and\ \citenamefont
  {Millan}}]{Humet1983}%
  \BibitemOpen
  \bibfield  {author} {\bibinfo {author} {\bibfnamefont {X.}~\bibnamefont
  {Aymerich-Humet}}, \bibinfo {author} {\bibfnamefont {F.}~\bibnamefont
  {Serra-Mestres}},\ and\ \bibinfo {author} {\bibfnamefont {J.}~\bibnamefont
  {Millan}},\ }\href {https://doi.org/10.1063/1.332276} {\bibfield  {journal}
  {\bibinfo  {journal} {Journal of Applied Physics}\ }\textbf {\bibinfo
  {volume} {54}},\ \bibinfo {pages} {2850} (\bibinfo {year}
  {1983})}\BibitemShut {NoStop}%
\bibitem [{\citenamefont {Guseinov}\ and\ \citenamefont
  {Mamedov}(2010)}]{Guseinov_2010}%
  \BibitemOpen
  \bibfield  {author} {\bibinfo {author} {\bibfnamefont {I.~I.}\ \bibnamefont
  {Guseinov}}\ and\ \bibinfo {author} {\bibfnamefont {B.~A.}\ \bibnamefont
  {Mamedov}},\ }\href {https://doi.org/10.1088/1674-1056/19/5/050501}
  {\bibfield  {journal} {\bibinfo  {journal} {Chinese Physics B}\ }\textbf
  {\bibinfo {volume} {19}},\ \bibinfo {pages} {050501} (\bibinfo {year}
  {2010})}\BibitemShut {NoStop}%
\bibitem [{\citenamefont {Fukushima}(2014)}]{fukushima2014417}%
  \BibitemOpen
  \bibfield  {author} {\bibinfo {author} {\bibfnamefont {T.}~\bibnamefont
  {Fukushima}},\ }\href
  {https://doi.org/https://doi.org/10.1016/j.amc.2014.02.053} {\bibfield
  {journal} {\bibinfo  {journal} {Applied Mathematics and Computation}\
  }\textbf {\bibinfo {volume} {234}},\ \bibinfo {pages} {417} (\bibinfo {year}
  {2014})}\BibitemShut {NoStop}%
\bibitem [{\citenamefont {Richter}\ \emph {et~al.}(2013)\citenamefont
  {Richter}, \citenamefont {Hermle},\ and\ \citenamefont
  {Glunz}}]{Richter_2013}%
  \BibitemOpen
  \bibfield  {author} {\bibinfo {author} {\bibfnamefont {A.}~\bibnamefont
  {Richter}}, \bibinfo {author} {\bibfnamefont {M.}~\bibnamefont {Hermle}},\
  and\ \bibinfo {author} {\bibfnamefont {S.~W.}\ \bibnamefont {Glunz}},\ }\href
  {https://doi.org/10.1109/JPHOTOV.2013.2270351} {\bibfield  {journal}
  {\bibinfo  {journal} {IEEE Journal of Photovoltaics}\ }\textbf {\bibinfo
  {volume} {3}},\ \bibinfo {pages} {1184} (\bibinfo {year} {2013})}\BibitemShut
  {NoStop}%
\bibitem [{\citenamefont {Daligou}\ \emph {et~al.}(2023)\citenamefont
  {Daligou}, \citenamefont {Attiaoui}, \citenamefont {Assali}, \citenamefont
  {Del~Vecchio},\ and\ \citenamefont {Moutanabbir}}]{RadLifetimeArxiv2023}%
  \BibitemOpen
  \bibfield  {author} {\bibinfo {author} {\bibfnamefont {G.}~\bibnamefont
  {Daligou}}, \bibinfo {author} {\bibfnamefont {A.}~\bibnamefont {Attiaoui}},
  \bibinfo {author} {\bibfnamefont {S.}~\bibnamefont {Assali}}, \bibinfo
  {author} {\bibfnamefont {P.}~\bibnamefont {Del~Vecchio}},\ and\ \bibinfo
  {author} {\bibfnamefont {O.}~\bibnamefont {Moutanabbir}},\ }\href
  {https://doi.org/10.48550/ARXIV.2302.02467} {\bibinfo {title} {Radiative
  carrier lifetime in ge$_{1-x}$sn$_x$ mid-infrared emitters}} (\bibinfo {year}
  {2023})\BibitemShut {NoStop}%
\bibitem [{\citenamefont {Streetman}\ and\ \citenamefont
  {Banerjee}(2000)}]{streetman2000}%
  \BibitemOpen
  \bibfield  {author} {\bibinfo {author} {\bibfnamefont {B.}~\bibnamefont
  {Streetman}}\ and\ \bibinfo {author} {\bibfnamefont {S.}~\bibnamefont
  {Banerjee}},\ }\href@noop {} {\emph {\bibinfo {title} {Solid State Electronic
  Devices}}},\ Prentice Hall series in solid state physical electronics\
  (\bibinfo  {publisher} {Prentice Hall},\ \bibinfo {year} {2000})\BibitemShut
  {NoStop}%
\bibitem [{\citenamefont {Hackenbuchner}(2002)}]{hackenbuchner2002}%
  \BibitemOpen
  \bibfield  {author} {\bibinfo {author} {\bibfnamefont {S.}~\bibnamefont
  {Hackenbuchner}},\ }\href@noop {} {\emph {\bibinfo {title} {Elektronische
  Struktur von Halbleiter-Nanobauelementen im thermodynamischen
  Nichtgleichgewicht}}},\ \bibinfo {edition} {1st}\ ed.,\ \bibinfo {series}
  {Ausgew{\"a}hlte Probleme der Halbleiterphysik und Technologie}\ No.~\bibinfo
  {number} {48}\ (\bibinfo  {publisher} {{Walter-Schottky-Inst}},\ \bibinfo
  {year} {2002})\BibitemShut {NoStop}%
\bibitem [{\citenamefont {Burger}\ \emph {et~al.}(2020)\citenamefont {Burger},
  \citenamefont {Sempere}, \citenamefont {{Roy-Layinde}},\ and\ \citenamefont
  {Lenert}}]{burgerPresentEfficienciesFuture2020a}%
  \BibitemOpen
  \bibfield  {author} {\bibinfo {author} {\bibfnamefont {T.}~\bibnamefont
  {Burger}}, \bibinfo {author} {\bibfnamefont {C.}~\bibnamefont {Sempere}},
  \bibinfo {author} {\bibfnamefont {B.}~\bibnamefont {{Roy-Layinde}}},\ and\
  \bibinfo {author} {\bibfnamefont {A.}~\bibnamefont {Lenert}},\ }\href
  {https://doi.org/10.1016/j.joule.2020.06.021} {\bibfield  {journal} {\bibinfo
   {journal} {Joule}\ }\textbf {\bibinfo {volume} {4}},\ \bibinfo {pages}
  {1660} (\bibinfo {year} {2020})}\BibitemShut {NoStop}%
\bibitem [{\citenamefont {Wernsman}\ \emph {et~al.}(2004)\citenamefont
  {Wernsman}, \citenamefont {Siergiej}, \citenamefont {Link}, \citenamefont
  {Mahorter}, \citenamefont {Palmisiano}, \citenamefont {Wehrer}, \citenamefont
  {Schultz}, \citenamefont {Schmuck}, \citenamefont {Messham}, \citenamefont
  {Murray}, \citenamefont {Murray}, \citenamefont {Newman}, \citenamefont
  {Taylor}, \citenamefont {DePoy},\ and\ \citenamefont
  {Rahmlow}}]{wernsmanGreater20Radiant2004}%
  \BibitemOpen
  \bibfield  {author} {\bibinfo {author} {\bibfnamefont {B.}~\bibnamefont
  {Wernsman}}, \bibinfo {author} {\bibfnamefont {R.}~\bibnamefont {Siergiej}},
  \bibinfo {author} {\bibfnamefont {S.}~\bibnamefont {Link}}, \bibinfo {author}
  {\bibfnamefont {R.}~\bibnamefont {Mahorter}}, \bibinfo {author}
  {\bibfnamefont {M.}~\bibnamefont {Palmisiano}}, \bibinfo {author}
  {\bibfnamefont {R.}~\bibnamefont {Wehrer}}, \bibinfo {author} {\bibfnamefont
  {R.}~\bibnamefont {Schultz}}, \bibinfo {author} {\bibfnamefont
  {G.}~\bibnamefont {Schmuck}}, \bibinfo {author} {\bibfnamefont
  {R.}~\bibnamefont {Messham}}, \bibinfo {author} {\bibfnamefont
  {S.}~\bibnamefont {Murray}}, \bibinfo {author} {\bibfnamefont
  {C.}~\bibnamefont {Murray}}, \bibinfo {author} {\bibfnamefont
  {F.}~\bibnamefont {Newman}}, \bibinfo {author} {\bibfnamefont
  {D.}~\bibnamefont {Taylor}}, \bibinfo {author} {\bibfnamefont
  {D.}~\bibnamefont {DePoy}},\ and\ \bibinfo {author} {\bibfnamefont
  {T.}~\bibnamefont {Rahmlow}},\ }\href
  {https://doi.org/10.1109/TED.2003.823247} {\bibfield  {journal} {\bibinfo
  {journal} {IEEE Transactions on Electron Devices}\ }\textbf {\bibinfo
  {volume} {51}},\ \bibinfo {pages} {512} (\bibinfo {year} {2004})}\BibitemShut
  {NoStop}%
\bibitem [{\citenamefont {Mauk}\ \emph {et~al.}(2003)\citenamefont {Mauk},
  \citenamefont {Sulima}, \citenamefont {Cox},\ and\ \citenamefont
  {Mueller}}]{maukLowbandgapEVInAsSbP2003}%
  \BibitemOpen
  \bibfield  {author} {\bibinfo {author} {\bibfnamefont {M.}~\bibnamefont
  {Mauk}}, \bibinfo {author} {\bibfnamefont {O.}~\bibnamefont {Sulima}},
  \bibinfo {author} {\bibfnamefont {J.}~\bibnamefont {Cox}},\ and\ \bibinfo
  {author} {\bibfnamefont {R.}~\bibnamefont {Mueller}},\ }in\ \href@noop {}
  {\emph {\bibinfo {booktitle} {Proceedings of 3rd {{World Conference on
  Photovoltaic Energy Conversion}}, 2003}}},\ Vol.~\bibinfo {volume} {1}\
  (\bibinfo {year} {2003})\ pp.\ \bibinfo {pages} {224--227 Vol.1}\BibitemShut
  {NoStop}%
\bibitem [{\citenamefont {Cheetham}\ \emph {et~al.}(2011)\citenamefont
  {Cheetham}, \citenamefont {Carrington}, \citenamefont {Cook},\ and\
  \citenamefont {Krier}}]{cheethamLowBandgapGaInAsSbP2011}%
  \BibitemOpen
  \bibfield  {author} {\bibinfo {author} {\bibfnamefont {K.~J.}\ \bibnamefont
  {Cheetham}}, \bibinfo {author} {\bibfnamefont {P.~J.}\ \bibnamefont
  {Carrington}}, \bibinfo {author} {\bibfnamefont {N.~B.}\ \bibnamefont
  {Cook}},\ and\ \bibinfo {author} {\bibfnamefont {A.}~\bibnamefont {Krier}},\
  }\href {https://doi.org/10.1016/j.solmat.2010.08.036} {\bibfield  {journal}
  {\bibinfo  {journal} {Solar Energy Materials and Solar Cells}\ }\textbf
  {\bibinfo {volume} {95}},\ \bibinfo {pages} {534} (\bibinfo {year}
  {2011})}\BibitemShut {NoStop}%
\bibitem [{\citenamefont {Huang}\ \emph {et~al.}(2020)\citenamefont {Huang},
  \citenamefont {Massengale}, \citenamefont {Lin}, \citenamefont {Li},
  \citenamefont {Yang}, \citenamefont {Mishima},\ and\ \citenamefont
  {Santos}}]{huangPerformanceAnalysisNarrowbandgap2020}%
  \BibitemOpen
  \bibfield  {author} {\bibinfo {author} {\bibfnamefont {W.}~\bibnamefont
  {Huang}}, \bibinfo {author} {\bibfnamefont {J.~A.}\ \bibnamefont
  {Massengale}}, \bibinfo {author} {\bibfnamefont {Y.}~\bibnamefont {Lin}},
  \bibinfo {author} {\bibfnamefont {L.}~\bibnamefont {Li}}, \bibinfo {author}
  {\bibfnamefont {R.~Q.}\ \bibnamefont {Yang}}, \bibinfo {author}
  {\bibfnamefont {T.~D.}\ \bibnamefont {Mishima}},\ and\ \bibinfo {author}
  {\bibfnamefont {M.~B.}\ \bibnamefont {Santos}},\ }\href
  {https://doi.org/10.1088/1361-6463/ab71b0} {\bibfield  {journal} {\bibinfo
  {journal} {Journal of Physics D: Applied Physics}\ }\textbf {\bibinfo
  {volume} {53}},\ \bibinfo {pages} {175104} (\bibinfo {year}
  {2020})}\BibitemShut {NoStop}%
\bibitem [{\citenamefont {Bose}\ \emph {et~al.}(2018)\citenamefont {Bose},
  \citenamefont {Cunha}, \citenamefont {Suresh}, \citenamefont {De~Wild},
  \citenamefont {Lopes}, \citenamefont {Barbosa}, \citenamefont {Silva},
  \citenamefont {Borme}, \citenamefont {Fernandes}, \citenamefont {Vermang},\
  and\ \citenamefont {Salom{\'e}}}]{bose2018}%
  \BibitemOpen
  \bibfield  {author} {\bibinfo {author} {\bibfnamefont {S.}~\bibnamefont
  {Bose}}, \bibinfo {author} {\bibfnamefont {J.~M.~V.}\ \bibnamefont {Cunha}},
  \bibinfo {author} {\bibfnamefont {S.}~\bibnamefont {Suresh}}, \bibinfo
  {author} {\bibfnamefont {J.}~\bibnamefont {De~Wild}}, \bibinfo {author}
  {\bibfnamefont {T.~S.}\ \bibnamefont {Lopes}}, \bibinfo {author}
  {\bibfnamefont {J.~R.~S.}\ \bibnamefont {Barbosa}}, \bibinfo {author}
  {\bibfnamefont {R.}~\bibnamefont {Silva}}, \bibinfo {author} {\bibfnamefont
  {J.}~\bibnamefont {Borme}}, \bibinfo {author} {\bibfnamefont {P.~A.}\
  \bibnamefont {Fernandes}}, \bibinfo {author} {\bibfnamefont {B.}~\bibnamefont
  {Vermang}},\ and\ \bibinfo {author} {\bibfnamefont {P.~M.~P.}\ \bibnamefont
  {Salom{\'e}}},\ }\href {https://doi.org/10.1002/solr.201800212} {\bibfield
  {journal} {\bibinfo  {journal} {Solar RRL}\ }\textbf {\bibinfo {volume}
  {2}},\ \bibinfo {pages} {1800212} (\bibinfo {year} {2018})}\BibitemShut
  {NoStop}%
\end{thebibliography}%
\bibliographystyle{apsrev4-2} % Tell bibtex which bibliography style to use

\end{document}